\documentclass[prl,twocolumn,superscriptaddress,showpacs]{revtex4-1}
\usepackage{amsmath,graphicx}
\usepackage[normalem]{ulem}
\usepackage{color}
\usepackage{bm}
\usepackage{epsfig}

\newcommand{\be}{\begin{equation}}
\newcommand{\ee}{\end{equation}}
\newcommand{\bea}{\begin{eqnarray}}
\newcommand{\eea}{\end{eqnarray}}
\newcommand{\ba}{\begin{eqnarray*}}
\newcommand{\ea}{\end{eqnarray*}}

\newcommand{\m}[1]{\mathcal{#1}}

\begin{document}

\title{Dissipation-Induced Superradiance in a Non-Markovian Open Dicke Model} 
\author{Orazio Scarlatella}
\author{Marco Schir\'o}
\affiliation{Institut de Physique Th\'{e}orique, Universit\'{e} Paris Saclay, CNRS, CEA, F-91191 Gif-sur-Yvette, France}
\date{\today}

\begin{abstract}
We consider the Dicke model, describing an ensemble of $N$ quantum spins interacting with a cavity field, and study how the coupling to a non-Markovian environment with power-law spectrum changes the physics of superradiant phase transition. Quite remarkably we find that dissipation can induce, rather than suppress, the ordered phase, a result which is in striking contrast with both thermal and Markovian quantum baths.  We interpret this dissipation-induced superradiance  as a genuine dissipative quantum phase transition that exists even at finite $N$ due to the coupling with the bath modes and whose nature and critical properties strongly depend on the spectral features of the non-Markovian environment.
\end{abstract}

\maketitle
\emph{Introduction - }Recent advances in quantum optics and quantum electronics have brought forth novel classes of hybrid systems where light and matter play equally important roles in emergent collective many body phenomena~\cite{AndrewNatPhys,CarusottoCiutiRMP13,RMP_Esslinger13,SchmidtKochAnnPhy13,LeHurReview16,NohAngelakisRepProgPhys2016,
HartmannJOpt2016}.
A crucial feature of these systems is their intrinsic dissipative nature: light-matter excitations are characterized by finite lifetime due to unavoidable losses, dephasing and decoherence processes originating from their coupling to external electromagnetic environments. This has stimulated a new wave of interest around open dissipative quantum many body systems at the interface between quantum optics and condensed matter physics. 


A paradigmatic example of collective light-matter phenomena is the Dicke model, describing $N$ quantum spins interacting with a photon field, with its associated superradiant (SR) phase transition~\cite{hepp_superradiant_1973,EmaryBrandesPRE03,garraway_dicke_2011}.  Stimulated by its experimental realization with ultra cold atoms in optical cavities~\cite{baumann_dicke_2010,Dicke_Esslinger_PRL11,brennecke_real-time_2013,baden_realization_2014,KlinderEtAlPNAS15} the dynamics of the open Dicke model in presence of a Markovian, memory-less, bath has been studied by a number of authors~\cite{HeppLieb73,dimer_proposed_2007,NagyEtAlPRL10,keeling_collective_2010,nagy_critical_2011,
oztop_excitations_2012,BhaseenEtAlPRA12,torre_keldysh_2013,kulkarni_cavity-mediated_2013,PiazzaStrackZwergerAnnPhys13,RylandsAndrei_arxiv14,LangPiazzaPRA16}, including photon losses and more recently spin dissipative processes~\cite{DallaTorreEtAl_arxiv16,GelhausenEtAl_arxiv16}. Quite generically (but see Ref.~\onlinecite{KirtonKeeling_arxiv16} for a recent interesting counter-example) the effect of losses is to shift the SR phase transition toward larger values of light-matter coupling, thus favouring the normal phase, much akin to a finite effective temperature, an analogy which has been put forward~\cite{torre_keldysh_2013} and shown to capture a number non trivial aspects of the Markovian open Dicke model, although some genuine new features remain.
 
An intriguing question which motivates this work is to understand the generality of this scenario for open-dissipative quantum systems and if one can conceive situations in which coupling to a quantum environment acts as a resource, rather than a limitation, for quantum state preparation and to engineer novel many body states. Such a \emph{dissipation engineering} is actively investigated in quantum optics~\cite{DiehlEtalNatPhys08,VerstraeteWolfCiracNatPhys09,Siddiqi_quantum_bath_engineering,AronKulkarniTureciPRA14,
HacohenEtAlPRL15,KimchiEtAlPRL16,HafeziEtAlPRB15}, where the idea is to design dissipation in such a way to reach a desired steady state. This is usually framed in the contex of Markovian baths, although the interest around memory effects and non-Markovianity in quantum optics is rapidly growing~\cite{RivasHuelgaPlenioRepProgPhys14,BreuerEtalRMP16}. In condensed matter physics, on the other hand, it is well known that quantum baths with rich low-frequency structure can mediate interactions and even drive phase transitions, the spin-boson problem being the most celebrated example~\cite{Leggett-RMP,Vojta_PhilMag06,LeHurReviewAnnPhys08}. 

The aim of this Letter is to bridge the gap between these two different approaches to quantum dissipative systems by studying the effect of a frequency dependent, non-Markovian quantum bath on the Dicke SR phase transition. We show that the phase diagram in the thermodynamic limit is qualitatively different from the Markovian case and highly non-thermal,  with coupling to the environment promoting, rather than suppressing, a dissipative SR phase. Quite interestingly we argue that this result is the mean-field limit of a genuine dissipative quantum phase transition persisting at finite $N$ for a thermodynamically large ohmic or sub-ohmic bath. Our results suggest that non-Markovian quantum baths can offer a way to engineer dissipative quantum states beyond the effective thermal picture, possibly even far from equilibrium when supplemented by external driving sources. 

\emph{The Dicke Model coupled to a Bath - } We start introducing the Dicke Hamiltonian,
\be \label{eq:DickeHam}
H_D=\omega_0\,a^{\dagger}a+\omega_q\,\sum_{i=1}^N\,S^z_i +\frac{2\lambda}{\sqrt{N}}\,\left(a+a^{\dagger}\right)\,\sum_{i=1}^{N}\,S^x_i
\ee
describing a set of $N$ spins $1/2$  $S_i^{\alpha=x,y,z}$, coupled to a single mode of the electromagnetic field with annihilation and creation operators $a,a^{\dagger}$. The frequencies of the photonic and spin modes read respectively as $\omega_0,\omega_q$ while the light-matter coupling is $\lambda$. The above Hamiltonian has a discrete $Z_2$ symmetry associated with a change of sign of both photon and spin operators, $(a,S^x)\rightarrow(-a,-S^x)$. This discrete symmetry can be spontaneously broken at zero temperature above a critical coupling $\lambda^0_c=\sqrt{\omega_0\omega_q}/2$ when the system enters a SR phase characterized by spontaneous polarizations, $\langle a+a^{\dagger}\rangle,\langle S^x_i\rangle\neq 0 $, as well as by a macroscopic occupation of the cavity mode, $\langle a^{\dagger}a\rangle\sim o(N)$. Upon heating, the superradiant phase gets destroyed above a critical temperature $T_c(\lambda)$ through a conventional Ising thermal phase transition~\cite{EmaryBrandesPRE03,SM_nonmarkovDicke}.
In order to study the effect of a non-Markovian bath on the problem, we couple the photon field to a structured electromagnetic environment described by a set of bosonic modes $c_k,c^{\dagger}_k$, such that the full Hamiltonian, including rotating ($a^{\dagger}c_k+hc$) and counter-rotating ($a^{\dagger}c^{\dagger}_k+hc$) terms, reads
\be\label{eqn:DissDicke}
H=H_D +  \sum_k \omega_k c_k^\dagger c_k + (a+a^\dagger )\sum_k g_k  (c_k + c_k^\dagger )
\ee
Crucial properties of the bath are encoded in its spectral function, $J_+(\omega)=\pi \sum_{k}g_k^2 \delta(\omega-\omega_k)$ which we will model in the following as $J_+(\omega)=\frac{\kappa}{2}\left(\omega/\omega_c\right)^s$, for $0<\omega<\omega_c$,  where $\kappa$ is the strength of the bath coupling and $\omega_c$ a cut-off. 

\emph{Solution for $N\rightarrow\infty$ -} To solve the model in presence of a generic bath we use the fact that the cavity field only couples to the total spin of the system, $S^{\alpha}=\sum_i S_i^{\alpha}$, which is conserved, i.e. $[H_D,S^2]=0$ with $S^{2}=\sum_{\alpha}\left(S^{\alpha}\right)^2$. If we restrict our attention to the sector with maximum spin, $S=N/2$, then the thermodynamic limit $N\rightarrow\infty$ coincides with the large spin limit in which the light-matter problem becomes harmonic~\cite{EmaryBrandesPRE03}. Indeed if we introduce a bosonic representation for the large spin, as $S^z=b^{\dagger}b-N/2$ and $S^+=b^{\dagger}\,\,\sqrt{N-b^{\dagger}b}$ the bosonized Dicke hamiltonian reads for $N\rightarrow\infty$
\be\label{eqn:bosons}
H_D=\omega_q\,b^{\dagger}b+\omega_0\,a^{\dagger}a+\lambda\,x\,X
\ee
where we have introduced the displacement fields $x=a+a^{\dagger}$, $X=b+b^{\dagger}$. The coupling to the environment, Eq.~(\ref{eqn:DissDicke}), can be accounted for exactly by integrating over the bath modes $c_k,c^{\dagger}_k$ in a Keldysh path integral to obtain an effective action for the $a,b$ subsystem~\cite{SM_nonmarkovDicke}. The effect of the bath on the spectral and dynamical properties of the system is encoded in the hybridization function 
$\Delta^R(\omega)=\Lambda(\omega)-iJ(\omega)$ with $\Lambda(\omega)=\mathcal{P}\int \frac{d\xi}{\pi} J_+(\xi)\frac{2\xi}{\omega^2-\xi^2}$ and $J(\omega)=\mbox{sign}(\omega)J_+(\vert\omega\vert)$. We assume an equilibrium zero temperature distribution of bath modes. The interplay between SR, non-equilibrium effects due to a drive~(see for example Ref.~\onlinecite{BastidasEtAlPRL12} for the closed system case) and dissipation due to a non-Markovian bath is an interesting issue  on which we will report elsewhere~\cite{ScarlatellaSchiro_inprepa}.

\emph{Normal Phase Instability - }To map the phase diagram of the open Dicke model we compute the susceptibility to a static coherent drive coupled to the $Z_2$ order parameter of the photon $x$, i.e. the retarded Green's function of the cavity field displacement,
\be\label{eqn:chiR}
\chi_a^R(\omega) = -i\int _0^{\infty}dt\,e^{-i\omega t} \langle [x(t),x(0)] \rangle 
\ee
While we focus on the cavity mode, we stress that a completely analogous result would have been obtained for the response function of the spin, $\chi_b^R(\omega)=-i\int _0^{\infty}dt\,e^{-i\omega t} \langle [X(t),X(0)] \rangle $, except at the special point $\lambda=0$ where the two subsystems decouple. After a simple calculation~\cite{SM_nonmarkovDicke} one can obtain the correlator~\eqref{eqn:chiR} which reads
\be\label{eqn:chiR_explicit}
\chi_a^R(\omega)=\frac{2\omega_0}{\omega^2-\omega_0^2-2\omega_0\left(\Delta^R(\omega)+\Sigma^R_{\lambda}(\omega)\right)} 
\ee
where the self-energy  $\Sigma^R(\omega) = \Delta^R(\omega)+\Sigma^R_{\lambda}(\omega)$ includes contributions from both the light matter coupling, i.e. $\Sigma^R_{\lambda}(\omega)=2\lambda^2\omega_q/\left(\omega^2-\omega_q^2\right)$, and the system-bath coupling $\Delta^R(\omega)$.  By taking the static limit $\omega\rightarrow0$ we find a diverging susceptibility $\chi^R(0)$, indicating an instability of the system toward a $Z_2$ symmetry-breaking, at a critical coupling $\lambda^{\infty}_c(\kappa)$
\be\label{eqn:lambda_mf}
\lambda^{\infty}_c(\kappa)=\sqrt{\tilde{\omega}_0(\kappa)\omega_q}/2
\ee
where $\tilde{\omega}_0(\kappa)=\omega_0+2\Lambda(0)=\omega_0-\frac{2\kappa}{\pi s}$ is the renormalized photon frequency, result of a bath-induced collective Lamb shift~\cite{FragnerScience08,ScullyPRL09,RohlsbergerScience10}.
\begin{figure}[t]
\begin{center}
\epsfig{figure=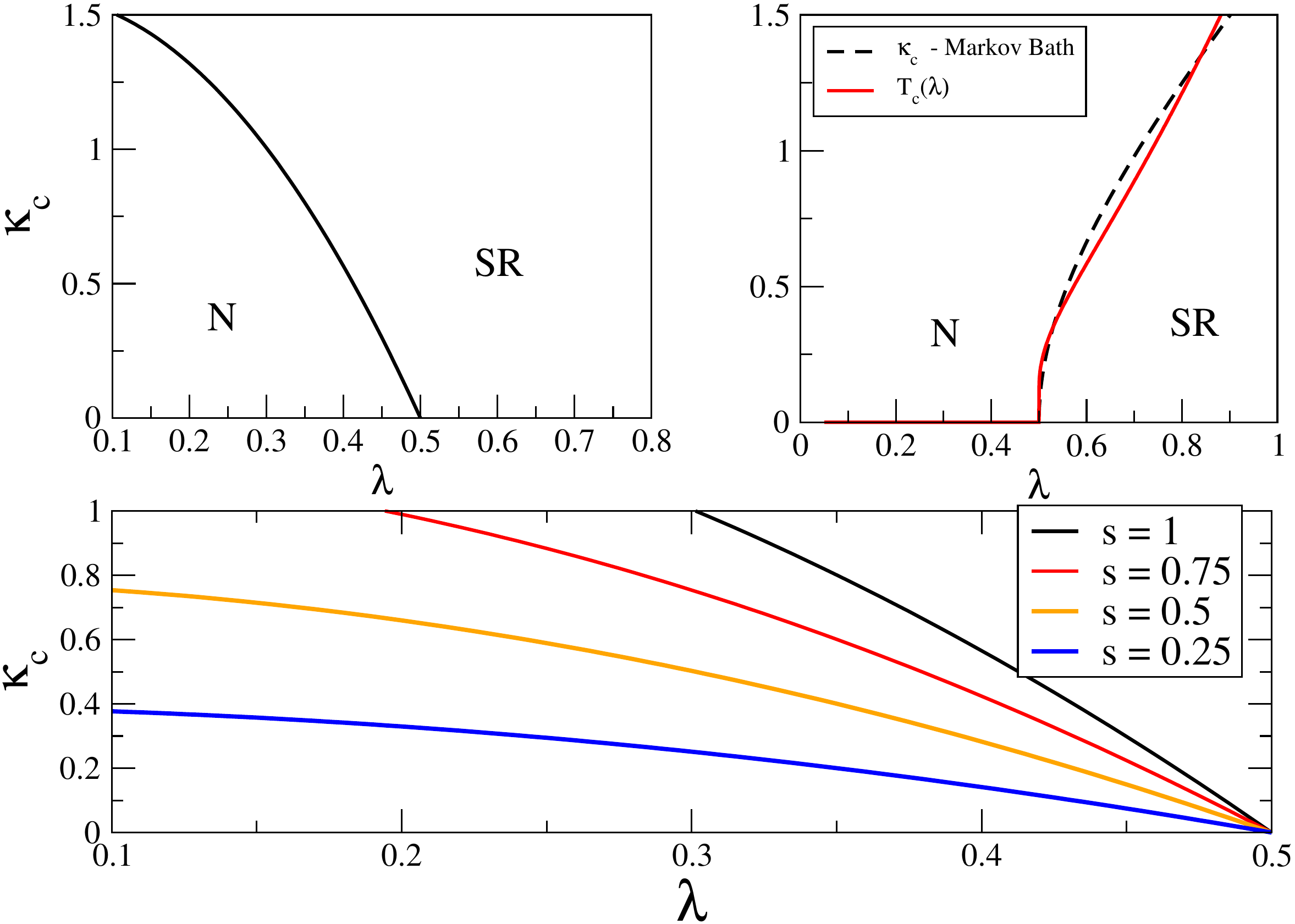,scale=0.36}
\caption{Non-Markovian Open Dicke Model in the thermodynamic limit. Top left panel: phase diagram for $s=1$, $\omega_0=\omega_q=1$. Top right panel: for comparison we plot the critical bath coupling in the Markovian case, $\kappa_c$  and the equilibrium critical temperature $T_c$ as a function of light-matter interaction $\lambda$. Bottom panel: critical bath coupling, $\kappa_c(\lambda)$ for different bath exponents $s$. }
\label{fig:fig1}
\end{center}
\end{figure}
We plot in figure \ref{fig:fig1} the phase boundary between normal (N) and SR phase, in the dissipation/light-matter coupling plane $(\kappa,\lambda)$. The critical coupling $\lambda^{\infty}_c(\kappa)$ reduces for $\kappa=0$, as expected, to the critical point for N-SR phase transition in the closed system. Quite importantly, we notice the phase boundary bends left-ward, namely the normal phase becomes unstable upon increasing the coupling to the bath beyond a certain threshold. In other words, non-Markovian dissipation induces a SR phase transition, an effect which is further enhanced upon reducing the bath exponent $s$, i.e. for a strongly sub-ohmic environment. Overall, this behavior is \textit{highly non-thermal}: as we show in figure~\ref{fig:fig1}, finite temperature effects are expected to destroy the SR phase above $T_c$, rather than promote it. Quite interestingly our phase boundary also qualitatively differs from the Markovian case (see fig.~\ref{fig:fig1} right panel) where photon losses with rate $\kappa$ are known~\cite{NagyEtAlPRL10,keeling_collective_2010} to \emph{increase} the critical point, $\lambda^M_c(\kappa)=\sqrt{\left(\omega_0^2+\kappa^2\right)\omega_q/4\omega_0}$~\cite{SM_nonmarkovDicke}. Before presenting a more general interpretation of our results it is useful to complete the picture and briefly discuss the properties of this dissipative SR phase.

\emph{Dissipative Superradiance - } Our treatment of the open Dicke model~(\ref{eqn:DissDicke}) has been confined so far to the normal phase and its instability toward $Z_2$ symmetry breaking. As for the isolated problem~\cite{EmaryBrandesPRE03}, in order to access the SR phase one needs to go beyond the Hamiltonian~(\ref{eqn:bosons}). This can be done in the present case by considering shifted operators for both the cavity, $a=\delta a+\sqrt{\alpha}$, the bosonized spin $b=\delta b-\sqrt{\beta}$ as well as the bath modes $c_k = \delta c_k - \sqrt{\gamma_k}$, and choosing the values of the c-numbers  $\alpha$, $\beta$ and $\{ \gamma_k \}$ in order to cancel the linear terms in the fluctuations $\delta a $, $\delta b$, $\delta c_k$ from Eq.~\eqref{eqn:DissDicke}. We can then compute the average value of the photon displacement operator which play the role of order parameter for the dissipative SR transition and reads
\begin{align}\label{eqn:x_SR}
\langle x \rangle &= 2\sqrt{\alpha} = \frac{\lambda}{\tilde{\omega}_0(\kappa)}\sqrt{4 N\left(1-\frac{\lambda^{\infty}_c(\kappa)^2}{\lambda^2}\right)} 
\end{align}
where the critical point $\lambda^{\infty}_c(\kappa)$ is the same obtained coming from the normal phase, Eq.~(\ref{eqn:lambda_mf}). Close to the critical point we have $\langle x \rangle \sim \left( \lambda-\lambda^{\infty}_c(\kappa)\right)^\frac{1}{2}$, a result fully consistent with the mean field nature of the $N=\infty$ transition (see below).  We see from Eq.~(\ref{eqn:x_SR}) that in the dissipative SR phase the cavity mode becomes coherent and macroscopically occupied, similarly to what happens in the closed system or open-Markovian case. One can then wonder what unique feature of the problem reflects the non-Markovian nature of the bath. In the rest of the paper we highlight the major and most striking one,  a complete analysis of static and dynamical properties of dissipative SR phase will be given elsewhere~\cite{ScarlatellaSchiro_inprepa}.

\begin{figure}[t]
\begin{center}
\epsfig{figure=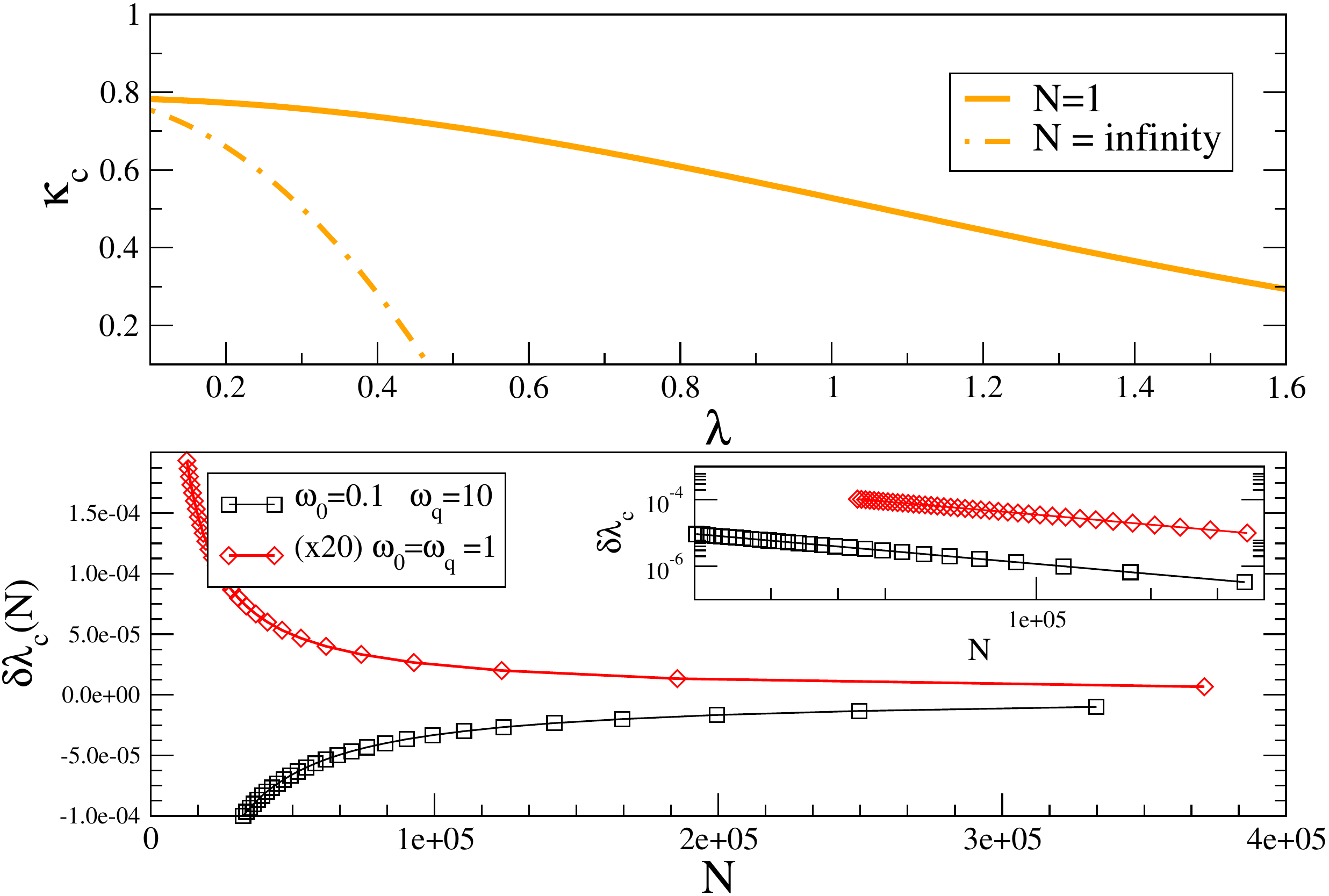,scale=0.35}
\caption{Top panel: Phase boundary of the effective spin-boson model for $N=1$ (see text): above a critical dissipation strenght the spin polarizes along the $x$ direction due to a bath-mediated exchange. This corresponds to a finite $N$ dissipative superradiance. Bottom panel: scaling of the deviation of the critical point from the mean field limit, $\delta \lambda_c(N,\kappa)=\lambda^N_c(\kappa)-\lambda^{\infty}_c(\kappa)\sim 1/N$. Parameters: $\omega_c=100$, $s=0.5$ $\kappa=0.5$ (black square) $\kappa=0.05$ (red diamonds).}
\label{fig:fig2}
\end{center}
\end{figure}

\emph{Effective Spin-Boson Model and Finite $N$ Transition - }  The results obtained so far are valid in the thermodynamic limit $N\rightarrow\infty$ when the large spin can be mapped onto an harmonic boson. In the rest of the paper we show that, however, these results should be seen as the mean-field limit of a genuine dissipative quantum phase transition that exists in the non-Markovian open Dicke model even at finite $N$ due to the coupling with the bath modes and whose nature and critical properties strongly depend on the spectral feauture of the environment. In order to uncover the non trivial finite $N$ aspect of the problem we proceed by integrating out all the bosonic modes, namely the bath and the cavity mode, to obtain an effective action for the spins degrees of freedom~\cite{Schiro_PRL12,SM_nonmarkovDicke} which reads 
 \be\label{eqn:Seff}
 \m{S}_{eff}= \frac{1}{N}\sum_{ij}\int d\tau d\tau' S^x_i(\tau)\,\m{J}_{eff}\left(\tau-\tau'\right)S^x_j(\tau')+\m{S}_{loc}
 \ee
where $\m{S}_{loc}$ is the local-in-time action of the quantum spin, whose explicit form is not needed in the following, while the effective, dynamical exchange $\m{J}_{eff}(\tau)$ is fully determined by the non-interacting propagator of the photon mode $\chi_{a0}^R(\omega)$ (see Eq.~(\ref{eqn:chiR_explicit}) for $\lambda=0$). To understand the physics of the finite $N$ non-Markovian open Dicke model it is instructive consider the case $N=1$, which reduces to the well known spin-boson model with a renormalized effective bath spectral function $\m{J}_{eff}(\omega)=-2\lambda^2 \mbox{Im}\chi^R_{a0}(\omega)\equiv 2\pi \alpha_{eff}\omega_c \left(\omega/\omega_c\right)^s\theta(\omega_c-\omega)$, with 
 \be\label{eqn:alpha_eff}
 \alpha_{eff}= \frac{\lambda^2\kappa}{\pi\omega_c\tilde{\omega}_0(\kappa)^2}
 \ee
for which many results are available~\cite{Leggett-RMP,LeHurReviewAnnPhys08}. In particular it is known that there is a zero temperature localization transition as a function of the coupling to the bath $\alpha_{eff}$, separating a phase where the two spin states $S^z=\pm1/2$ are delocalized, i.e. $\langle S^x\rangle=0$ for  $\alpha_{eff}< \alpha_c$ from a localized phase with broken symmetry $\langle S^x\rangle \neq0$ for  $\alpha_{eff}>\alpha_c$, due to the long-ranged nature of the retarded interaction $\m{J}_{eff}\left(\tau-\tau'\right)$. The critical dissipation strength $\alpha_c$ is essentially known exactly for ohmic bath, $\alpha_c=1+o(\omega_q/\omega_c)$ while in the subohmic case  a recently proposed variational wave function~\cite{Plenio_prl11,BeraEtAlPRB14} has been shown to qualitatively capture the physics of this transition and to give $\alpha_c = \sin\pi s e^{-s/2}/2\pi(1-s)\,\left(\omega_q/\omega_c\right)^{1-s}$.
Using these results and Eq.(\ref{eqn:alpha_eff}) we can extract the critical line for the $N=1$ non-Markovian Dicke model, that that we plot in the $(\kappa,\lambda)$ plane in figure~\ref{fig:fig2} for $s=0.5$. Upon increasing the coupling to the bath the normal (delocalized) phase becomes unstable and the quantum spin gets localized. While Eq.~(\ref{eqn:Seff}) describes an effective theory for the spin, we stress that the photon mode still remains in the picture and becomes coherent in the broken symmetry phase.

Next we discuss what happens upon increasing $N$ and how the mean field result $\lambda_c^{\infty}(\kappa)$ is recovered in this spin-boson picture. 
To address this point it is useful to notice that the effective action in Eq.~(\ref{eqn:Seff}) contains two kind of contributions: (i) a term purely local in space, for $i=j$, corresponding to a set of $N$ independent spin-boson models whose coupling to the bath is \emph{suppressed by a factor $1/N$} and (ii) a non-local term coupling all-to-all spin-boson models at different sites $i\neq j$. Upon increasing the number of spins $N$ the first contribution is suppressed by the factor $1/N$, while the second one has a non vanishing thermodynamic limit and eventually dominates. Indeed if we take only the short-range (in time) part of the non-local exchange and we write it as $\m{J}_{eff}(\tau-\tau')\sim \bar{\m{J}}_{eff}\delta(\tau-\tau')$, with  $\bar{\m{J}}_{eff}=\int d\tau \m{J}_{eff}(\tau)=-\frac{4\lambda^2}{\tilde{\omega}_0(\kappa)}$ we obtain a static action corresponding to a fully connected Ising model in a transverse field,
\be\label{eqn:Ising}
H_{eff} = \frac{\bar{\m{J}}_{eff}}{N}\sum_{ij}S^x_i S^x_j+\omega_q\sum_i S^z 
\ee 
which has a quantum phase transition where the $Z_2$ breaks and the spin align along the $x$ direction, $\langle S^x\rangle\neq 0$. This transition happens when $\bar{\m{J}}_{eff}=-\omega_q$, a condition which immediately gives the mean field phase boundary $\lambda^{\infty}_{c}(\kappa)$ in Eq.~(\ref{eqn:lambda_mf}). 
The above analysis suggests to define an $N$-dependent effective dissipation strenght $\alpha_{eff}^N(\lambda,\kappa)$ scaling as
\be\label{eqn:alpha_effN}
\alpha^N_{eff}(\lambda,\kappa)=\alpha^{\infty}_{eff}(\lambda,\kappa)+ \alpha_{eff}(\lambda,\kappa)/N
\ee    
with  $\alpha^{\infty}_{eff}=\lambda^2s/\tilde{\omega}_0(\kappa)\omega_c$ and $\alpha_{eff}=\lambda^2\kappa/\pi\tilde{\omega}_0^2(\kappa)\omega_c$~\cite{SM_nonmarkovDicke}. At finite $N$ the action~(\ref{eqn:Seff}) describes a multi-spin-boson model~\cite{VorrathBrandesPRL05,AndersNJP08,WinterRiegerPRB14,HenrietLeHurPRB16} whose critical properties have been the subject of recent numerical investigations indicating that a zero temperature localization-delocalization transition survives at any $N$ for ohmic and sub-ohmic baths ($s\le 1$) with robust scaling with $N$ of the critical dissipation strength, $\alpha^N_c(\omega_q,s)=\alpha^{\infty}_c(\omega_q,s)+\delta\alpha(\omega_q)/N$, where $\alpha^{\infty}_c=s\omega_q/4\omega_c$  
 while $\delta\alpha$ does not depend on the bath exponent~\cite{WinterRiegerPRB14}.  
 Within our mapping~(\ref{eqn:alpha_effN}) this would immediately give $\lambda^N_c(\kappa)=\lambda^{\infty}_c(\kappa)+\delta\lambda/N$, with deviations $\delta\lambda$ positive or negative depending on $\omega_0,\omega_q$~\cite{SM_nonmarkovDicke}.  To verify this picture we include perturbatively finite-$N$ corrections to the mean field Dicke Hamiltonian~(\ref{eqn:bosons}), which amounts to supplement the quadratic bosonic theory by an interaction term $H_{int}=-(\lambda/2N)b^{\dagger}x_a x_b b$. By computing the retarded Green's function for the boson $b$ displacement, $\chi_b^R(t)$, a proxy to the spin-spin correlation, we extract the phase boundary at finite $N$. In figure ~\ref{fig:fig2} we plot the deviation of the critical point from the mean field $N=\infty$ result, $\delta\lambda_c(N)=\lambda^N_c-\lambda^{\infty}_c$, which confirms the predicted scaling. We stress that the finite-N dissipation induced SR transition only exists for $s\leq 1$ (while its mean field version, Eq. (\ref{eqn:lambda_mf}), for $s>0$) and its critical properties depend strongly on the bath spectral function exponent $s$, much like the spin-boson transition. In particular we expect the universality class of the finite $N$ transition to be Kosterlitz-Thouless for $s=1$ while continuous with s-dependent exponents for $s<1$ ~\cite{Vojta_PhilMag06,WinterRiegerPRB14}. As a result we expect an interesting RG flow at finite $N$ whose investigation we leave for future studies. 

\emph{Discussion - }  We have already emphasized the qualitative difference, in thermodynamic limit, between our phase diagram (fig.~\ref{fig:fig1}) and the one obtained with a markovian or thermal bath. Such a difference is even more striking at finite $N$, where the dissipative SR phase transition of figure~\ref{fig:fig2} could not arise in presence of a fast decaying, memory-less, Markovian environment. On the other hand our results substantially differ also from those obtained recently for a seemingly related non-Markovian open Dicke model~\cite{NagyDomokosPRL15,NagyDomokos_arxiv16}. There however, one of the two bosonic modes remains coupled to a Markovian bath, while the second one hybridizes with a frequency-structured environment. This crucial difference might explain the results obtained in the thermodynamic limit, with the SR critical point unchanged by the colored bath which instead affects some property of the system at the transition, most notably the occupation of bosonic mode. A detailed comparison including finite $N$ effects in presence of two baths will be presented elsewhere~\cite{ScarlatellaSchiro_inprepa}.

Finally, it is interesting to comment on possible extensions of this work. The finite $N$ spin-boson problem~(\ref{eqn:Seff}) can be amenable to further analytical approaches to study its properties non-perturbatively in $1/N$, a relevant issue to understand the approach to the mean field phase boundary at small dissipation, $\kappa\rightarrow0$, where our lowest order perturbation theory becomes singular. The addition of an external drive is also particular appealing for investigations of effective thermalization in non-Markovian environments. In this respect it is useful to mention that the properties of driven, large $S$, spin-boson models are largely unexplored and much of the understanding of its phase transition is based on a quantum-classical mapping for imaginary time evolution, which calls for further work. Finally, a natural question is how to engineer non-Markovian dissipation that allows to experimentally access this interesting dissipative quantum phase transition and its interplay with nonequilibrium effects. In this respect it is worth noticing many recent progresses in controlling quantum environments using different platforms such as cavity and circuit QED~\cite{LiuEtAlNatPhys11,SundaresanEtAlPRX15,HaeberleinEtAl_arxiv15,LiuHouckNatPhys16}, which also offer the natural setting to realize Dicke-like quantum light-matter models~\cite{FinkEtAlPRL09,GrezesEtAlPRX14,KakuyanagiEtAlPRL16}.

To conclude, in this paper we have studied the effect of a non-Markovian quantum environment on the paradigmatic light-matter Dicke phase transition. We have seen that quite differently from the memory-less case, dissipation induced by the bath with abundant low-frequency modes is able to drive a genuine quantum phase transition toward a dissipative superradiant phase, whose mean field limit we have studied in detail. Interestingly, at finite $N$ the transition remains sharp for a thermodynamically large ohmic or subohmic bath with a non trivial critical behavior that we linked to the physics of a generalized spin-boson problem.

\emph{Acknownledgements - } We acknowledge discussions with N. Andrei, J. Bhaseen, S. Gopalakrishnan, D. Huse, A. Mitra and H. Tureci. This work was supported  by the CNRS through the PICS-USA-147504 and by a grant "Investissements d'Avenir" from LabEx PALM (ANR-10-LABX-0039-PALM).


\begin{thebibliography}{64}%
\makeatletter
\providecommand \@ifxundefined [1]{%
 \@ifx{#1\undefined}
}%
\providecommand \@ifnum [1]{%
 \ifnum #1\expandafter \@firstoftwo
 \else \expandafter \@secondoftwo
 \fi
}%
\providecommand \@ifx [1]{%
 \ifx #1\expandafter \@firstoftwo
 \else \expandafter \@secondoftwo
 \fi
}%
\providecommand \natexlab [1]{#1}%
\providecommand \enquote  [1]{``#1''}%
\providecommand \bibnamefont  [1]{#1}%
\providecommand \bibfnamefont [1]{#1}%
\providecommand \citenamefont [1]{#1}%
\providecommand \href@noop [0]{\@secondoftwo}%
\providecommand \href [0]{\begingroup \@sanitize@url \@href}%
\providecommand \@href[1]{\@@startlink{#1}\@@href}%
\providecommand \@@href[1]{\endgroup#1\@@endlink}%
\providecommand \@sanitize@url [0]{\catcode `\\12\catcode `\$12\catcode
  `\&12\catcode `\#12\catcode `\^12\catcode `\_12\catcode `\%12\relax}%
\providecommand \@@startlink[1]{}%
\providecommand \@@endlink[0]{}%
\providecommand \url  [0]{\begingroup\@sanitize@url \@url }%
\providecommand \@url [1]{\endgroup\@href {#1}{\urlprefix }}%
\providecommand \urlprefix  [0]{URL }%
\providecommand \Eprint [0]{\href }%
\providecommand \doibase [0]{http://dx.doi.org/}%
\providecommand \selectlanguage [0]{\@gobble}%
\providecommand \bibinfo  [0]{\@secondoftwo}%
\providecommand \bibfield  [0]{\@secondoftwo}%
\providecommand \translation [1]{[#1]}%
\providecommand \BibitemOpen [0]{}%
\providecommand \bibitemStop [0]{}%
\providecommand \bibitemNoStop [0]{.\EOS\space}%
\providecommand \EOS [0]{\spacefactor3000\relax}%
\providecommand \BibitemShut  [1]{\csname bibitem#1\endcsname}%
\let\auto@bib@innerbib\@empty
\bibitem [{\citenamefont {Houck}\ \emph {et~al.}(2012)\citenamefont {Houck},
  \citenamefont {Tureci},\ and\ \citenamefont {Koch}}]{AndrewNatPhys}%
  \BibitemOpen
  \bibfield  {author} {\bibinfo {author} {\bibfnamefont {A.~A.}\ \bibnamefont
  {Houck}}, \bibinfo {author} {\bibfnamefont {H.~E.}\ \bibnamefont {Tureci}}, \
  and\ \bibinfo {author} {\bibfnamefont {J.}~\bibnamefont {Koch}},\ }\href@noop
  {} {\bibfield  {journal} {\bibinfo  {journal} {Nature Physics}\ }\textbf
  {\bibinfo {volume} {8}} (\bibinfo {year} {2012})}\BibitemShut {NoStop}%
\bibitem [{\citenamefont {Carusotto}\ and\ \citenamefont
  {Ciuti}(2013)}]{CarusottoCiutiRMP13}%
  \BibitemOpen
  \bibfield  {author} {\bibinfo {author} {\bibfnamefont {I.}~\bibnamefont
  {Carusotto}}\ and\ \bibinfo {author} {\bibfnamefont {C.}~\bibnamefont
  {Ciuti}},\ }\href@noop {} {\bibfield  {journal} {\bibinfo  {journal} {Rev.
  Mod. Phys.}\ }\textbf {\bibinfo {volume} {85}},\ \bibinfo {pages} {299}
  (\bibinfo {year} {2013})}\BibitemShut {NoStop}%
\bibitem [{\citenamefont {Ritsch}\ \emph {et~al.}(2013)\citenamefont {Ritsch},
  \citenamefont {Domokos}, \citenamefont {Brennecke},\ and\ \citenamefont
  {Esslinger}}]{RMP_Esslinger13}%
  \BibitemOpen
  \bibfield  {author} {\bibinfo {author} {\bibfnamefont {H.}~\bibnamefont
  {Ritsch}}, \bibinfo {author} {\bibfnamefont {P.}~\bibnamefont {Domokos}},
  \bibinfo {author} {\bibfnamefont {F.}~\bibnamefont {Brennecke}}, \ and\
  \bibinfo {author} {\bibfnamefont {T.}~\bibnamefont {Esslinger}},\ }\href@noop
  {} {\bibfield  {journal} {\bibinfo  {journal} {Rev. Mod. Phys.}\ }\textbf
  {\bibinfo {volume} {85}},\ \bibinfo {pages} {553} (\bibinfo {year}
  {2013})}\BibitemShut {NoStop}%
\bibitem [{\citenamefont {Schmidt}\ and\ \citenamefont
  {Koch}(2013)}]{SchmidtKochAnnPhy13}%
  \BibitemOpen
  \bibfield  {author} {\bibinfo {author} {\bibfnamefont {S.}~\bibnamefont
  {Schmidt}}\ and\ \bibinfo {author} {\bibfnamefont {J.}~\bibnamefont {Koch}},\
  }\href {\doibase 10.1002/andp.201200261} {\bibfield  {journal} {\bibinfo
  {journal} {Annalen der Physik}\ }\textbf {\bibinfo {volume} {525}},\ \bibinfo
  {pages} {395} (\bibinfo {year} {2013})}\BibitemShut {NoStop}%
\bibitem [{\citenamefont {Hur}\ \emph {et~al.}(2016)\citenamefont {Hur},
  \citenamefont {Henriet}, \citenamefont {Petrescu}, \citenamefont {Plekhanov},
  \citenamefont {Roux},\ and\ \citenamefont {Schir\'o}}]{LeHurReview16}%
  \BibitemOpen
  \bibfield  {author} {\bibinfo {author} {\bibfnamefont {K.~L.}\ \bibnamefont
  {Hur}}, \bibinfo {author} {\bibfnamefont {L.}~\bibnamefont {Henriet}},
  \bibinfo {author} {\bibfnamefont {A.}~\bibnamefont {Petrescu}}, \bibinfo
  {author} {\bibfnamefont {K.}~\bibnamefont {Plekhanov}}, \bibinfo {author}
  {\bibfnamefont {G.}~\bibnamefont {Roux}}, \ and\ \bibinfo {author}
  {\bibfnamefont {M.}~\bibnamefont {Schir\'o}},\ }\href@noop {} {\bibfield
  {journal} {\bibinfo  {journal} {Comptes Rendus Physique}\ }\textbf {\bibinfo
  {volume} {17}},\ \bibinfo {pages} {808 } (\bibinfo {year}
  {2016})}\BibitemShut {NoStop}%
\bibitem [{\citenamefont {Noh}\ and\ \citenamefont
  {Angelakis}(2017)}]{NohAngelakisRepProgPhys2016}%
  \BibitemOpen
  \bibfield  {author} {\bibinfo {author} {\bibfnamefont {C.}~\bibnamefont
  {Noh}}\ and\ \bibinfo {author} {\bibfnamefont {D.~G.}\ \bibnamefont
  {Angelakis}},\ }\href@noop {} {\bibfield  {journal} {\bibinfo  {journal}
  {Reports on Progress in Physics}\ }\textbf {\bibinfo {volume} {80}},\
  \bibinfo {pages} {016401} (\bibinfo {year} {2017})}\BibitemShut {NoStop}%
\bibitem [{\citenamefont {Hartmann}(2016)}]{HartmannJOpt2016}%
  \BibitemOpen
  \bibfield  {author} {\bibinfo {author} {\bibfnamefont {M.~J.}\ \bibnamefont
  {Hartmann}},\ }\href@noop {} {\bibfield  {journal} {\bibinfo  {journal}
  {Journal of Optics}\ }\textbf {\bibinfo {volume} {18}},\ \bibinfo {pages}
  {104005} (\bibinfo {year} {2016})}\BibitemShut {NoStop}%
\bibitem [{\citenamefont {Hepp}\ and\ \citenamefont
  {Lieb}(1973{\natexlab{a}})}]{hepp_superradiant_1973}%
  \BibitemOpen
  \bibfield  {author} {\bibinfo {author} {\bibfnamefont {K.}~\bibnamefont
  {Hepp}}\ and\ \bibinfo {author} {\bibfnamefont {E.~H.}\ \bibnamefont
  {Lieb}},\ }\href {\doibase 10.1016/0003-4916(73)90039-0} {\bibfield
  {journal} {\bibinfo  {journal} {Annals of Physics}\ }\textbf {\bibinfo
  {volume} {76}},\ \bibinfo {pages} {360} (\bibinfo {year}
  {1973}{\natexlab{a}})}\BibitemShut {NoStop}%
\bibitem [{\citenamefont {Emary}\ and\ \citenamefont
  {Brandes}(2003)}]{EmaryBrandesPRE03}%
  \BibitemOpen
  \bibfield  {author} {\bibinfo {author} {\bibfnamefont {C.}~\bibnamefont
  {Emary}}\ and\ \bibinfo {author} {\bibfnamefont {T.}~\bibnamefont
  {Brandes}},\ }\href@noop {} {\bibfield  {journal} {\bibinfo  {journal} {Phys.
  Rev. E}\ }\textbf {\bibinfo {volume} {67}},\ \bibinfo {pages} {066203}
  (\bibinfo {year} {2003})}\BibitemShut {NoStop}%
\bibitem [{\citenamefont {Garraway}(2011)}]{garraway_dicke_2011}%
  \BibitemOpen
  \bibfield  {author} {\bibinfo {author} {\bibfnamefont {B.~M.}\ \bibnamefont
  {Garraway}},\ }\href {\doibase 10.1098/rsta.2010.0333} {\bibfield  {journal}
  {\bibinfo  {journal} {Phil. Trans. Roy. Soc. A.}\ }\textbf {\bibinfo {volume}
  {369}},\ \bibinfo {pages} {1137} (\bibinfo {year} {2011})}\BibitemShut
  {NoStop}%
\bibitem [{\citenamefont {Baumann}\ \emph {et~al.}(2010)\citenamefont
  {Baumann}, \citenamefont {Guerlin}, \citenamefont {Brennecke},\ and\
  \citenamefont {Esslinger}}]{baumann_dicke_2010}%
  \BibitemOpen
  \bibfield  {author} {\bibinfo {author} {\bibfnamefont {K.}~\bibnamefont
  {Baumann}}, \bibinfo {author} {\bibfnamefont {C.}~\bibnamefont {Guerlin}},
  \bibinfo {author} {\bibfnamefont {F.}~\bibnamefont {Brennecke}}, \ and\
  \bibinfo {author} {\bibfnamefont {T.}~\bibnamefont {Esslinger}},\ }\href@noop
  {} {\bibfield  {journal} {\bibinfo  {journal} {Nature}\ }\textbf {\bibinfo
  {volume} {464}},\ \bibinfo {pages} {1301} (\bibinfo {year}
  {2010})}\BibitemShut {NoStop}%
\bibitem [{\citenamefont {Baumann}\ \emph {et~al.}(2011)\citenamefont
  {Baumann}, \citenamefont {Mottl}, \citenamefont {Brennecke},\ and\
  \citenamefont {Esslinger}}]{Dicke_Esslinger_PRL11}%
  \BibitemOpen
  \bibfield  {author} {\bibinfo {author} {\bibfnamefont {K.}~\bibnamefont
  {Baumann}}, \bibinfo {author} {\bibfnamefont {R.}~\bibnamefont {Mottl}},
  \bibinfo {author} {\bibfnamefont {F.}~\bibnamefont {Brennecke}}, \ and\
  \bibinfo {author} {\bibfnamefont {T.}~\bibnamefont {Esslinger}},\ }\href
  {\doibase 10.1103/PhysRevLett.107.140402} {\bibfield  {journal} {\bibinfo
  {journal} {Phys. Rev. Lett.}\ }\textbf {\bibinfo {volume} {107}},\ \bibinfo
  {pages} {140402} (\bibinfo {year} {2011})}\BibitemShut {NoStop}%
\bibitem [{\citenamefont {Brennecke}\ \emph {et~al.}(2013)\citenamefont
  {Brennecke}, \citenamefont {Mottl}, \citenamefont {Baumann}, \citenamefont
  {Landig}, \citenamefont {Donner},\ and\ \citenamefont
  {Esslinger}}]{brennecke_real-time_2013}%
  \BibitemOpen
  \bibfield  {author} {\bibinfo {author} {\bibfnamefont {F.}~\bibnamefont
  {Brennecke}}, \bibinfo {author} {\bibfnamefont {R.}~\bibnamefont {Mottl}},
  \bibinfo {author} {\bibfnamefont {K.}~\bibnamefont {Baumann}}, \bibinfo
  {author} {\bibfnamefont {R.}~\bibnamefont {Landig}}, \bibinfo {author}
  {\bibfnamefont {T.}~\bibnamefont {Donner}}, \ and\ \bibinfo {author}
  {\bibfnamefont {T.}~\bibnamefont {Esslinger}},\ }\href {\doibase
  10.1073/pnas.1306993110} {\bibfield  {journal} {\bibinfo  {journal} {Proc.
  Nat. Acad. Sci. USA}\ }\textbf {\bibinfo {volume} {110}},\ \bibinfo {pages}
  {11763} (\bibinfo {year} {2013})}\BibitemShut {NoStop}%
\bibitem [{\citenamefont {Baden}\ \emph {et~al.}(2014)\citenamefont {Baden},
  \citenamefont {Arnold}, \citenamefont {Grimsmo}, \citenamefont {Parkins},\
  and\ \citenamefont {Barrett}}]{baden_realization_2014}%
  \BibitemOpen
  \bibfield  {author} {\bibinfo {author} {\bibfnamefont {M.~P.}\ \bibnamefont
  {Baden}}, \bibinfo {author} {\bibfnamefont {K.~J.}\ \bibnamefont {Arnold}},
  \bibinfo {author} {\bibfnamefont {A.~L.}\ \bibnamefont {Grimsmo}}, \bibinfo
  {author} {\bibfnamefont {S.}~\bibnamefont {Parkins}}, \ and\ \bibinfo
  {author} {\bibfnamefont {M.~D.}\ \bibnamefont {Barrett}},\ }\href {\doibase
  10.1103/PhysRevLett.113.020408} {\bibfield  {journal} {\bibinfo  {journal}
  {Phys. Rev. Lett.}\ }\textbf {\bibinfo {volume} {113}},\ \bibinfo {pages}
  {020408} (\bibinfo {year} {2014})}\BibitemShut {NoStop}%
\bibitem [{\citenamefont {Klinder}\ \emph {et~al.}(2015)\citenamefont
  {Klinder}, \citenamefont {Keßler}, \citenamefont {Wolke}, \citenamefont
  {Mathey},\ and\ \citenamefont {Hemmerich}}]{KlinderEtAlPNAS15}%
  \BibitemOpen
  \bibfield  {author} {\bibinfo {author} {\bibfnamefont {J.}~\bibnamefont
  {Klinder}}, \bibinfo {author} {\bibfnamefont {H.}~\bibnamefont {Keßler}},
  \bibinfo {author} {\bibfnamefont {M.}~\bibnamefont {Wolke}}, \bibinfo
  {author} {\bibfnamefont {L.}~\bibnamefont {Mathey}}, \ and\ \bibinfo {author}
  {\bibfnamefont {A.}~\bibnamefont {Hemmerich}},\ }\href@noop {} {\bibfield
  {journal} {\bibinfo  {journal} {Proceedings of the National Academy of
  Sciences}\ }\textbf {\bibinfo {volume} {112}},\ \bibinfo {pages} {3290}
  (\bibinfo {year} {2015})}\BibitemShut {NoStop}%
\bibitem [{\citenamefont {Hepp}\ and\ \citenamefont
  {Lieb}(1973{\natexlab{b}})}]{HeppLieb73}%
  \BibitemOpen
  \bibfield  {author} {\bibinfo {author} {\bibfnamefont {K.}~\bibnamefont
  {Hepp}}\ and\ \bibinfo {author} {\bibfnamefont {E.~H.}\ \bibnamefont
  {Lieb}},\ }\href@noop {} {\bibfield  {journal} {\bibinfo  {journal} {Helv.
  Phys. Acta}\ }\textbf {\bibinfo {volume} {46}},\ \bibinfo {pages} {576}
  (\bibinfo {year} {1973}{\natexlab{b}})}\BibitemShut {NoStop}%
\bibitem [{\citenamefont {Dimer}\ \emph {et~al.}(2007)\citenamefont {Dimer},
  \citenamefont {Estienne}, \citenamefont {Parkins},\ and\ \citenamefont
  {Carmichael}}]{dimer_proposed_2007}%
  \BibitemOpen
  \bibfield  {author} {\bibinfo {author} {\bibfnamefont {F.}~\bibnamefont
  {Dimer}}, \bibinfo {author} {\bibfnamefont {B.}~\bibnamefont {Estienne}},
  \bibinfo {author} {\bibfnamefont {A.~S.}\ \bibnamefont {Parkins}}, \ and\
  \bibinfo {author} {\bibfnamefont {H.~J.}\ \bibnamefont {Carmichael}},\ }\href
  {\doibase 10.1103/PhysRevA.75.013804} {\bibfield  {journal} {\bibinfo
  {journal} {Phys. Rev. A}\ }\textbf {\bibinfo {volume} {75}},\ \bibinfo
  {pages} {013804} (\bibinfo {year} {2007})}\BibitemShut {NoStop}%
\bibitem [{\citenamefont {Nagy}\ \emph {et~al.}(2010)\citenamefont {Nagy},
  \citenamefont {K\'onya}, \citenamefont {Szirmai},\ and\ \citenamefont
  {Domokos}}]{NagyEtAlPRL10}%
  \BibitemOpen
  \bibfield  {author} {\bibinfo {author} {\bibfnamefont {D.}~\bibnamefont
  {Nagy}}, \bibinfo {author} {\bibfnamefont {G.}~\bibnamefont {K\'onya}},
  \bibinfo {author} {\bibfnamefont {G.}~\bibnamefont {Szirmai}}, \ and\
  \bibinfo {author} {\bibfnamefont {P.}~\bibnamefont {Domokos}},\ }\href@noop
  {} {\bibfield  {journal} {\bibinfo  {journal} {Phys. Rev. Lett.}\ }\textbf
  {\bibinfo {volume} {104}},\ \bibinfo {pages} {130401} (\bibinfo {year}
  {2010})}\BibitemShut {NoStop}%
\bibitem [{\citenamefont {Keeling}\ \emph {et~al.}(2010)\citenamefont
  {Keeling}, \citenamefont {Bhaseen},\ and\ \citenamefont
  {Simons}}]{keeling_collective_2010}%
  \BibitemOpen
  \bibfield  {author} {\bibinfo {author} {\bibfnamefont {J.}~\bibnamefont
  {Keeling}}, \bibinfo {author} {\bibfnamefont {M.~J.}\ \bibnamefont
  {Bhaseen}}, \ and\ \bibinfo {author} {\bibfnamefont {B.~D.}\ \bibnamefont
  {Simons}},\ }\href {\doibase 10.1103/PhysRevLett.105.043001} {\bibfield
  {journal} {\bibinfo  {journal} {Phys. Rev. Lett.}\ }\textbf {\bibinfo
  {volume} {105}},\ \bibinfo {pages} {043001} (\bibinfo {year}
  {2010})}\BibitemShut {NoStop}%
\bibitem [{\citenamefont {Nagy}\ \emph {et~al.}(2011)\citenamefont {Nagy},
  \citenamefont {Szirmai},\ and\ \citenamefont {Domokos}}]{nagy_critical_2011}%
  \BibitemOpen
  \bibfield  {author} {\bibinfo {author} {\bibfnamefont {D.}~\bibnamefont
  {Nagy}}, \bibinfo {author} {\bibfnamefont {G.}~\bibnamefont {Szirmai}}, \
  and\ \bibinfo {author} {\bibfnamefont {P.}~\bibnamefont {Domokos}},\ }\href
  {\doibase 10.1103/PhysRevA.84.043637} {\bibfield  {journal} {\bibinfo
  {journal} {Phys. Rev. A}\ }\textbf {\bibinfo {volume} {84}},\ \bibinfo
  {pages} {043637} (\bibinfo {year} {2011})}\BibitemShut {NoStop}%
\bibitem [{\citenamefont {{\"O}ztop}\ \emph {et~al.}(2012)\citenamefont
  {{\"O}ztop}, \citenamefont {Bordyuh}, \citenamefont {M{\"u}stecapl{\i}o{\u
  g}lu},\ and\ \citenamefont {T{\"u}reci}}]{oztop_excitations_2012}%
  \BibitemOpen
  \bibfield  {author} {\bibinfo {author} {\bibfnamefont {B.}~\bibnamefont
  {{\"O}ztop}}, \bibinfo {author} {\bibfnamefont {M.}~\bibnamefont {Bordyuh}},
  \bibinfo {author} {\bibfnamefont {{\"O}.~E.}\ \bibnamefont
  {M{\"u}stecapl{\i}o{\u g}lu}}, \ and\ \bibinfo {author} {\bibfnamefont
  {H.~E.}\ \bibnamefont {T{\"u}reci}},\ }\href {\doibase
  10.1088/1367-2630/14/8/085011} {\bibfield  {journal} {\bibinfo  {journal}
  {New J. Phys.}\ }\textbf {\bibinfo {volume} {14}},\ \bibinfo {pages} {085011}
  (\bibinfo {year} {2012})}\BibitemShut {NoStop}%
\bibitem [{\citenamefont {Bhaseen}\ \emph {et~al.}(2012)\citenamefont
  {Bhaseen}, \citenamefont {Mayoh}, \citenamefont {Simons},\ and\ \citenamefont
  {Keeling}}]{BhaseenEtAlPRA12}%
  \BibitemOpen
  \bibfield  {author} {\bibinfo {author} {\bibfnamefont {M.~J.}\ \bibnamefont
  {Bhaseen}}, \bibinfo {author} {\bibfnamefont {J.}~\bibnamefont {Mayoh}},
  \bibinfo {author} {\bibfnamefont {B.~D.}\ \bibnamefont {Simons}}, \ and\
  \bibinfo {author} {\bibfnamefont {J.}~\bibnamefont {Keeling}},\ }\href
  {\doibase 10.1103/PhysRevA.85.013817} {\bibfield  {journal} {\bibinfo
  {journal} {Phys. Rev. A}\ }\textbf {\bibinfo {volume} {85}},\ \bibinfo
  {pages} {013817} (\bibinfo {year} {2012})}\BibitemShut {NoStop}%
\bibitem [{\citenamefont {Torre}\ \emph {et~al.}(2013)\citenamefont {Torre},
  \citenamefont {Diehl}, \citenamefont {Lukin}, \citenamefont {Sachdev},\ and\
  \citenamefont {Strack}}]{torre_keldysh_2013}%
  \BibitemOpen
  \bibfield  {author} {\bibinfo {author} {\bibfnamefont {E.~G.~D.}\
  \bibnamefont {Torre}}, \bibinfo {author} {\bibfnamefont {S.}~\bibnamefont
  {Diehl}}, \bibinfo {author} {\bibfnamefont {M.~D.}\ \bibnamefont {Lukin}},
  \bibinfo {author} {\bibfnamefont {S.}~\bibnamefont {Sachdev}}, \ and\
  \bibinfo {author} {\bibfnamefont {P.}~\bibnamefont {Strack}},\ }\href
  {\doibase 10.1103/PhysRevA.87.023831} {\bibfield  {journal} {\bibinfo
  {journal} {Phys. Rev. A}\ }\textbf {\bibinfo {volume} {87}},\ \bibinfo
  {pages} {023831} (\bibinfo {year} {2013})}\BibitemShut {NoStop}%
\bibitem [{\citenamefont {Kulkarni}\ \emph {et~al.}(2013)\citenamefont
  {Kulkarni}, \citenamefont {{\"O}ztop},\ and\ \citenamefont
  {T{\"u}reci}}]{kulkarni_cavity-mediated_2013}%
  \BibitemOpen
  \bibfield  {author} {\bibinfo {author} {\bibfnamefont {M.}~\bibnamefont
  {Kulkarni}}, \bibinfo {author} {\bibfnamefont {B.}~\bibnamefont {{\"O}ztop}},
  \ and\ \bibinfo {author} {\bibfnamefont {H.~E.}\ \bibnamefont {T{\"u}reci}},\
  }\href {\doibase 10.1103/PhysRevLett.111.220408} {\bibfield  {journal}
  {\bibinfo  {journal} {Phys. Rev. Lett.}\ }\textbf {\bibinfo {volume} {111}},\
  \bibinfo {pages} {220408} (\bibinfo {year} {2013})}\BibitemShut {NoStop}%
\bibitem [{\citenamefont {Piazza}\ \emph {et~al.}(2013)\citenamefont {Piazza},
  \citenamefont {Strack},\ and\ \citenamefont
  {Zwerger}}]{PiazzaStrackZwergerAnnPhys13}%
  \BibitemOpen
  \bibfield  {author} {\bibinfo {author} {\bibfnamefont {F.}~\bibnamefont
  {Piazza}}, \bibinfo {author} {\bibfnamefont {P.}~\bibnamefont {Strack}}, \
  and\ \bibinfo {author} {\bibfnamefont {W.}~\bibnamefont {Zwerger}},\ }\href
  {\doibase http://dx.doi.org/10.1016/j.aop.2013.08.015} {\bibfield  {journal}
  {\bibinfo  {journal} {Annals of Physics}\ }\textbf {\bibinfo {volume}
  {339}},\ \bibinfo {pages} {135 } (\bibinfo {year} {2013})}\BibitemShut
  {NoStop}%
\bibitem [{\citenamefont {{Rylands}}\ and\ \citenamefont
  {{Andrei}}(2014)}]{RylandsAndrei_arxiv14}%
  \BibitemOpen
  \bibfield  {author} {\bibinfo {author} {\bibfnamefont {C.}~\bibnamefont
  {{Rylands}}}\ and\ \bibinfo {author} {\bibfnamefont {N.}~\bibnamefont
  {{Andrei}}},\ }\href@noop {} {\bibfield  {journal} {\bibinfo  {journal}
  {ArXiv e-prints}\ } (\bibinfo {year} {2014})},\ \Eprint
  {http://arxiv.org/abs/1408.3652} {arXiv:1408.3652 [physics.optics]}
  \BibitemShut {NoStop}%
\bibitem [{\citenamefont {Lang}\ and\ \citenamefont
  {Piazza}(2016)}]{LangPiazzaPRA16}%
  \BibitemOpen
  \bibfield  {author} {\bibinfo {author} {\bibfnamefont {J.}~\bibnamefont
  {Lang}}\ and\ \bibinfo {author} {\bibfnamefont {F.}~\bibnamefont {Piazza}},\
  }\href@noop {} {\bibfield  {journal} {\bibinfo  {journal} {Phys. Rev. A}\
  }\textbf {\bibinfo {volume} {94}},\ \bibinfo {pages} {033628} (\bibinfo
  {year} {2016})}\BibitemShut {NoStop}%
\bibitem [{\citenamefont {{Dalla Torre}}\ \emph {et~al.}(2016)\citenamefont
  {{Dalla Torre}}, \citenamefont {{Shchadilova}}, \citenamefont {{Wilner}},
  \citenamefont {{Lukin}},\ and\ \citenamefont
  {{Demler}}}]{DallaTorreEtAl_arxiv16}%
  \BibitemOpen
  \bibfield  {author} {\bibinfo {author} {\bibfnamefont {E.~G.}\ \bibnamefont
  {{Dalla Torre}}}, \bibinfo {author} {\bibfnamefont {Y.}~\bibnamefont
  {{Shchadilova}}}, \bibinfo {author} {\bibfnamefont {E.~Y.}\ \bibnamefont
  {{Wilner}}}, \bibinfo {author} {\bibfnamefont {M.~D.}\ \bibnamefont
  {{Lukin}}}, \ and\ \bibinfo {author} {\bibfnamefont {E.}~\bibnamefont
  {{Demler}}},\ }\href@noop {} {\bibfield  {journal} {\bibinfo  {journal}
  {ArXiv e-prints}\ } (\bibinfo {year} {2016})},\ \Eprint
  {http://arxiv.org/abs/1608.06293} {arXiv:1608.06293 [quant-ph]} \BibitemShut
  {NoStop}%
\bibitem [{\citenamefont {{Gelhausen}}\ \emph {et~al.}(2016)\citenamefont
  {{Gelhausen}}, \citenamefont {{Buchhold}},\ and\ \citenamefont
  {{Strack}}}]{GelhausenEtAl_arxiv16}%
  \BibitemOpen
  \bibfield  {author} {\bibinfo {author} {\bibfnamefont {J.}~\bibnamefont
  {{Gelhausen}}}, \bibinfo {author} {\bibfnamefont {M.}~\bibnamefont
  {{Buchhold}}}, \ and\ \bibinfo {author} {\bibfnamefont {P.}~\bibnamefont
  {{Strack}}},\ }\href@noop {} {\bibfield  {journal} {\bibinfo  {journal}
  {ArXiv e-prints}\ } (\bibinfo {year} {2016})},\ \Eprint
  {http://arxiv.org/abs/1605.07637} {arXiv:1605.07637 [cond-mat.quant-gas]}
  \BibitemShut {NoStop}%
\bibitem [{\citenamefont {{Kirton}}\ and\ \citenamefont
  {{Keeling}}(2016)}]{KirtonKeeling_arxiv16}%
  \BibitemOpen
  \bibfield  {author} {\bibinfo {author} {\bibfnamefont {P.}~\bibnamefont
  {{Kirton}}}\ and\ \bibinfo {author} {\bibfnamefont {J.}~\bibnamefont
  {{Keeling}}},\ }\href@noop {} {\bibfield  {journal} {\bibinfo  {journal}
  {ArXiv e-prints}\ } (\bibinfo {year} {2016})},\ \Eprint
  {http://arxiv.org/abs/1611.03342} {arXiv:1611.03342 [cond-mat.quant-gas]}
  \BibitemShut {NoStop}%
\bibitem [{\citenamefont {Diehl}\ \emph {et~al.}(2008)\citenamefont {Diehl},
  \citenamefont {Micheli}, \citenamefont {Kantian}, \citenamefont {Kraus},
  \citenamefont {Buchler},\ and\ \citenamefont {Zoller}}]{DiehlEtalNatPhys08}%
  \BibitemOpen
  \bibfield  {author} {\bibinfo {author} {\bibfnamefont {S.}~\bibnamefont
  {Diehl}}, \bibinfo {author} {\bibfnamefont {A.}~\bibnamefont {Micheli}},
  \bibinfo {author} {\bibfnamefont {A.}~\bibnamefont {Kantian}}, \bibinfo
  {author} {\bibfnamefont {B.}~\bibnamefont {Kraus}}, \bibinfo {author}
  {\bibfnamefont {H.~P.}\ \bibnamefont {Buchler}}, \ and\ \bibinfo {author}
  {\bibfnamefont {P.}~\bibnamefont {Zoller}},\ }\href
  {http://dx.doi.org/10.1038/nphys1073} {\bibfield  {journal} {\bibinfo
  {journal} {Nat Phys}\ }\textbf {\bibinfo {volume} {4}},\ \bibinfo {pages}
  {878} (\bibinfo {year} {2008})}\BibitemShut {NoStop}%
\bibitem [{\citenamefont {Verstraete}\ \emph {et~al.}(2009)\citenamefont
  {Verstraete}, \citenamefont {Wolf},\ and\ \citenamefont
  {Ignacio~Cirac}}]{VerstraeteWolfCiracNatPhys09}%
  \BibitemOpen
  \bibfield  {author} {\bibinfo {author} {\bibfnamefont {F.}~\bibnamefont
  {Verstraete}}, \bibinfo {author} {\bibfnamefont {M.~M.}\ \bibnamefont
  {Wolf}}, \ and\ \bibinfo {author} {\bibfnamefont {J.}~\bibnamefont
  {Ignacio~Cirac}},\ }\href@noop {} {\bibfield  {journal} {\bibinfo  {journal}
  {Nat Phys}\ }\textbf {\bibinfo {volume} {5}},\ \bibinfo {pages} {633}
  (\bibinfo {year} {2009})}\BibitemShut {NoStop}%
\bibitem [{\citenamefont {Murch}\ \emph {et~al.}(2012)\citenamefont {Murch},
  \citenamefont {Vool}, \citenamefont {Zhou}, \citenamefont {Weber},
  \citenamefont {Girvin},\ and\ \citenamefont
  {Siddiqi}}]{Siddiqi_quantum_bath_engineering}%
  \BibitemOpen
  \bibfield  {author} {\bibinfo {author} {\bibfnamefont {K.~W.}\ \bibnamefont
  {Murch}}, \bibinfo {author} {\bibfnamefont {U.}~\bibnamefont {Vool}},
  \bibinfo {author} {\bibfnamefont {D.}~\bibnamefont {Zhou}}, \bibinfo {author}
  {\bibfnamefont {S.~J.}\ \bibnamefont {Weber}}, \bibinfo {author}
  {\bibfnamefont {S.~M.}\ \bibnamefont {Girvin}}, \ and\ \bibinfo {author}
  {\bibfnamefont {I.}~\bibnamefont {Siddiqi}},\ }\href@noop {} {\bibfield
  {journal} {\bibinfo  {journal} {Phys. Rev. Lett.}\ }\textbf {\bibinfo
  {volume} {109}},\ \bibinfo {pages} {183602} (\bibinfo {year}
  {2012})}\BibitemShut {NoStop}%
\bibitem [{\citenamefont {Aron}\ \emph {et~al.}(2014)\citenamefont {Aron},
  \citenamefont {Kulkarni},\ and\ \citenamefont
  {T\"ureci}}]{AronKulkarniTureciPRA14}%
  \BibitemOpen
  \bibfield  {author} {\bibinfo {author} {\bibfnamefont {C.}~\bibnamefont
  {Aron}}, \bibinfo {author} {\bibfnamefont {M.}~\bibnamefont {Kulkarni}}, \
  and\ \bibinfo {author} {\bibfnamefont {H.~E.}\ \bibnamefont {T\"ureci}},\
  }\href {\doibase 10.1103/PhysRevA.90.062305} {\bibfield  {journal} {\bibinfo
  {journal} {Phys. Rev. A}\ }\textbf {\bibinfo {volume} {90}},\ \bibinfo
  {pages} {062305} (\bibinfo {year} {2014})}\BibitemShut {NoStop}%
\bibitem [{\citenamefont {Hacohen-Gourgy}\ \emph {et~al.}(2015)\citenamefont
  {Hacohen-Gourgy}, \citenamefont {Ramasesh}, \citenamefont {De~Grandi},
  \citenamefont {Siddiqi},\ and\ \citenamefont {Girvin}}]{HacohenEtAlPRL15}%
  \BibitemOpen
  \bibfield  {author} {\bibinfo {author} {\bibfnamefont {S.}~\bibnamefont
  {Hacohen-Gourgy}}, \bibinfo {author} {\bibfnamefont {V.~V.}\ \bibnamefont
  {Ramasesh}}, \bibinfo {author} {\bibfnamefont {C.}~\bibnamefont {De~Grandi}},
  \bibinfo {author} {\bibfnamefont {I.}~\bibnamefont {Siddiqi}}, \ and\
  \bibinfo {author} {\bibfnamefont {S.~M.}\ \bibnamefont {Girvin}},\ }\href
  {\doibase 10.1103/PhysRevLett.115.240501} {\bibfield  {journal} {\bibinfo
  {journal} {Phys. Rev. Lett.}\ }\textbf {\bibinfo {volume} {115}},\ \bibinfo
  {pages} {240501} (\bibinfo {year} {2015})}\BibitemShut {NoStop}%
\bibitem [{\citenamefont {Kimchi-Schwartz}\ \emph {et~al.}(2016)\citenamefont
  {Kimchi-Schwartz}, \citenamefont {Martin}, \citenamefont {Flurin},
  \citenamefont {Aron}, \citenamefont {Kulkarni}, \citenamefont {Tureci},\ and\
  \citenamefont {Siddiqi}}]{KimchiEtAlPRL16}%
  \BibitemOpen
  \bibfield  {author} {\bibinfo {author} {\bibfnamefont {M.~E.}\ \bibnamefont
  {Kimchi-Schwartz}}, \bibinfo {author} {\bibfnamefont {L.}~\bibnamefont
  {Martin}}, \bibinfo {author} {\bibfnamefont {E.}~\bibnamefont {Flurin}},
  \bibinfo {author} {\bibfnamefont {C.}~\bibnamefont {Aron}}, \bibinfo {author}
  {\bibfnamefont {M.}~\bibnamefont {Kulkarni}}, \bibinfo {author}
  {\bibfnamefont {H.~E.}\ \bibnamefont {Tureci}}, \ and\ \bibinfo {author}
  {\bibfnamefont {I.}~\bibnamefont {Siddiqi}},\ }\href {\doibase
  10.1103/PhysRevLett.116.240503} {\bibfield  {journal} {\bibinfo  {journal}
  {Phys. Rev. Lett.}\ }\textbf {\bibinfo {volume} {116}},\ \bibinfo {pages}
  {240503} (\bibinfo {year} {2016})}\BibitemShut {NoStop}%
\bibitem [{\citenamefont {Hafezi}\ \emph {et~al.}(2015)\citenamefont {Hafezi},
  \citenamefont {Adhikari},\ and\ \citenamefont {Taylor}}]{HafeziEtAlPRB15}%
  \BibitemOpen
  \bibfield  {author} {\bibinfo {author} {\bibfnamefont {M.}~\bibnamefont
  {Hafezi}}, \bibinfo {author} {\bibfnamefont {P.}~\bibnamefont {Adhikari}}, \
  and\ \bibinfo {author} {\bibfnamefont {J.~M.}\ \bibnamefont {Taylor}},\
  }\href {\doibase 10.1103/PhysRevB.92.174305} {\bibfield  {journal} {\bibinfo
  {journal} {Phys. Rev. B}\ }\textbf {\bibinfo {volume} {92}},\ \bibinfo
  {pages} {174305} (\bibinfo {year} {2015})}\BibitemShut {NoStop}%
\bibitem [{\citenamefont {Ángel Rivas}\ \emph {et~al.}(2014)\citenamefont
  {Ángel Rivas}, \citenamefont {Huelga},\ and\ \citenamefont
  {Plenio}}]{RivasHuelgaPlenioRepProgPhys14}%
  \BibitemOpen
  \bibfield  {author} {\bibinfo {author} {\bibnamefont {Ángel Rivas}},
  \bibinfo {author} {\bibfnamefont {S.~F.}\ \bibnamefont {Huelga}}, \ and\
  \bibinfo {author} {\bibfnamefont {M.~B.}\ \bibnamefont {Plenio}},\ }\href
  {http://stacks.iop.org/0034-4885/77/i=9/a=094001} {\bibfield  {journal}
  {\bibinfo  {journal} {Reports on Progress in Physics}\ }\textbf {\bibinfo
  {volume} {77}},\ \bibinfo {pages} {094001} (\bibinfo {year}
  {2014})}\BibitemShut {NoStop}%
\bibitem [{\citenamefont {Breuer}\ \emph {et~al.}(2016)\citenamefont {Breuer},
  \citenamefont {Laine}, \citenamefont {Piilo},\ and\ \citenamefont
  {Vacchini}}]{BreuerEtalRMP16}%
  \BibitemOpen
  \bibfield  {author} {\bibinfo {author} {\bibfnamefont {H.-P.}\ \bibnamefont
  {Breuer}}, \bibinfo {author} {\bibfnamefont {E.-M.}\ \bibnamefont {Laine}},
  \bibinfo {author} {\bibfnamefont {J.}~\bibnamefont {Piilo}}, \ and\ \bibinfo
  {author} {\bibfnamefont {B.}~\bibnamefont {Vacchini}},\ }\href {\doibase
  10.1103/RevModPhys.88.021002} {\bibfield  {journal} {\bibinfo  {journal}
  {Rev. Mod. Phys.}\ }\textbf {\bibinfo {volume} {88}},\ \bibinfo {pages}
  {021002} (\bibinfo {year} {2016})}\BibitemShut {NoStop}%
\bibitem [{\citenamefont {Leggett}\ \emph {et~al.}(1987)\citenamefont
  {Leggett}, \citenamefont {Chakravarty}, \citenamefont {Dorsey}, \citenamefont
  {Fisher}, \citenamefont {Garg},\ and\ \citenamefont {Zwerger}}]{Leggett-RMP}%
  \BibitemOpen
  \bibfield  {author} {\bibinfo {author} {\bibfnamefont {A.~J.}\ \bibnamefont
  {Leggett}}, \bibinfo {author} {\bibfnamefont {S.}~\bibnamefont
  {Chakravarty}}, \bibinfo {author} {\bibfnamefont {A.~T.}\ \bibnamefont
  {Dorsey}}, \bibinfo {author} {\bibfnamefont {M.~P.~A.}\ \bibnamefont
  {Fisher}}, \bibinfo {author} {\bibfnamefont {A.}~\bibnamefont {Garg}}, \ and\
  \bibinfo {author} {\bibfnamefont {W.}~\bibnamefont {Zwerger}},\ }\href
  {\doibase 10.1103/RevModPhys.59.1} {\bibfield  {journal} {\bibinfo  {journal}
  {Rev. Mod. Phys.}\ }\textbf {\bibinfo {volume} {59}},\ \bibinfo {pages} {1}
  (\bibinfo {year} {1987})}\BibitemShut {NoStop}%
\bibitem [{\citenamefont {Vojta}(2006)}]{Vojta_PhilMag06}%
  \BibitemOpen
  \bibfield  {author} {\bibinfo {author} {\bibfnamefont {M.}~\bibnamefont
  {Vojta}},\ }\href {\doibase 10.1080/14786430500070396} {\bibfield  {journal}
  {\bibinfo  {journal} {Philosophical Magazine}\ }\textbf {\bibinfo {volume}
  {86}},\ \bibinfo {pages} {1807} (\bibinfo {year} {2006})}\BibitemShut
  {NoStop}%
\bibitem [{\citenamefont {Hur}(2008)}]{LeHurReviewAnnPhys08}%
  \BibitemOpen
  \bibfield  {author} {\bibinfo {author} {\bibfnamefont {K.~L.}\ \bibnamefont
  {Hur}},\ }\href@noop {} {\bibfield  {journal} {\bibinfo  {journal} {Annals of
  Physics}\ }\textbf {\bibinfo {volume} {323}},\ \bibinfo {pages} {2208 }
  (\bibinfo {year} {2008})}\BibitemShut {NoStop}%
\bibitem [{SM_()}]{SM_nonmarkovDicke}%
  \BibitemOpen
  \href@noop {} {\bibinfo  {journal} {See supplementary material for
  further details on finite temperature and Markovian-bath phase diagram, Keldysh calculation of photon susceptibility and spin-boson analysis}\ }\BibitemShut {NoStop}%
\bibitem [{\citenamefont {Bastidas}\ \emph {et~al.}(2012)\citenamefont
  {Bastidas}, \citenamefont {Emary}, \citenamefont {Regler},\ and\
  \citenamefont {Brandes}}]{BastidasEtAlPRL12}%
  \BibitemOpen
\bibfield  {journal} {  }\bibfield  {author} {\bibinfo {author} {\bibfnamefont
  {V.~M.}\ \bibnamefont {Bastidas}}, \bibinfo {author} {\bibfnamefont
  {C.}~\bibnamefont {Emary}}, \bibinfo {author} {\bibfnamefont
  {B.}~\bibnamefont {Regler}}, \ and\ \bibinfo {author} {\bibfnamefont
  {T.}~\bibnamefont {Brandes}},\ }\href@noop {} {\bibfield  {journal} {\bibinfo
   {journal} {Phys. Rev. Lett.}\ }\textbf {\bibinfo {volume} {108}},\ \bibinfo
  {pages} {043003} (\bibinfo {year} {2012})}\BibitemShut {NoStop}%
\bibitem [{\citenamefont {Scarlatella}\ and\ \citenamefont
  {Schir\'o}(2017)}]{ScarlatellaSchiro_inprepa}%
  \BibitemOpen
  \bibfield  {author} {\bibinfo {author} {\bibfnamefont {O.}~\bibnamefont
  {Scarlatella}}\ and\ \bibinfo {author} {\bibfnamefont {M.}~\bibnamefont
  {Schir\'o}},\ }\href@noop {} {\bibfield  {journal} {\bibinfo  {journal} {in
  preparation}\ } (\bibinfo {year} {2017})}\BibitemShut {NoStop}%
\bibitem [{\citenamefont {Fragner}\ \emph {et~al.}(2008)\citenamefont
  {Fragner}, \citenamefont {G\"oppl}, \citenamefont {Fink}, \citenamefont
  {Baur}, \citenamefont {Bianchetti}, \citenamefont {Leek}, \citenamefont
  {Blais},\ and\ \citenamefont {Wallraff}}]{FragnerScience08}%
  \BibitemOpen
  \bibfield  {author} {\bibinfo {author} {\bibfnamefont {A.}~\bibnamefont
  {Fragner}}, \bibinfo {author} {\bibfnamefont {M.}~\bibnamefont {G\"oppl}},
  \bibinfo {author} {\bibfnamefont {J.~M.}\ \bibnamefont {Fink}}, \bibinfo
  {author} {\bibfnamefont {M.}~\bibnamefont {Baur}}, \bibinfo {author}
  {\bibfnamefont {R.}~\bibnamefont {Bianchetti}}, \bibinfo {author}
  {\bibfnamefont {P.~J.}\ \bibnamefont {Leek}}, \bibinfo {author}
  {\bibfnamefont {A.}~\bibnamefont {Blais}}, \ and\ \bibinfo {author}
  {\bibfnamefont {A.}~\bibnamefont {Wallraff}},\ }\href {\doibase
  10.1126/science.1164482} {\bibfield  {journal} {\bibinfo  {journal}
  {Science}\ }\textbf {\bibinfo {volume} {322}},\ \bibinfo {pages} {1357}
  (\bibinfo {year} {2008})}\BibitemShut {NoStop}%
\bibitem [{\citenamefont {Scully}(2009)}]{ScullyPRL09}%
  \BibitemOpen
  \bibfield  {author} {\bibinfo {author} {\bibfnamefont {M.~O.}\ \bibnamefont
  {Scully}},\ }\href@noop {} {\bibfield  {journal} {\bibinfo  {journal} {Phys.
  Rev. Lett.}\ }\textbf {\bibinfo {volume} {102}},\ \bibinfo {pages} {143601}
  (\bibinfo {year} {2009})}\BibitemShut {NoStop}%
\bibitem [{\citenamefont {R\"ohlsberger}\ \emph {et~al.}(2010)\citenamefont
  {R\"ohlsberger}, \citenamefont {Schlage}, \citenamefont {Sahoo},
  \citenamefont {Couet},\ and\ \citenamefont
  {R\"uffer}}]{RohlsbergerScience10}%
  \BibitemOpen
  \bibfield  {author} {\bibinfo {author} {\bibfnamefont {R.}~\bibnamefont
  {R\"ohlsberger}}, \bibinfo {author} {\bibfnamefont {K.}~\bibnamefont
  {Schlage}}, \bibinfo {author} {\bibfnamefont {B.}~\bibnamefont {Sahoo}},
  \bibinfo {author} {\bibfnamefont {S.}~\bibnamefont {Couet}}, \ and\ \bibinfo
  {author} {\bibfnamefont {R.}~\bibnamefont {R\"uffer}},\ }\href {\doibase
  10.1126/science.1187770} {\bibfield  {journal} {\bibinfo  {journal}
  {Science}\ }\textbf {\bibinfo {volume} {328}},\ \bibinfo {pages} {1248}
  (\bibinfo {year} {2010})}\BibitemShut {NoStop}%
\bibitem [{\citenamefont {Schir\'o}\ \emph {et~al.}(2012)\citenamefont
  {Schir\'o}, \citenamefont {Bordyuh}, \citenamefont {\"Oztop},\ and\
  \citenamefont {T\"ureci}}]{Schiro_PRL12}%
  \BibitemOpen
  \bibfield  {author} {\bibinfo {author} {\bibfnamefont {M.}~\bibnamefont
  {Schir\'o}}, \bibinfo {author} {\bibfnamefont {M.}~\bibnamefont {Bordyuh}},
  \bibinfo {author} {\bibfnamefont {B.}~\bibnamefont {\"Oztop}}, \ and\
  \bibinfo {author} {\bibfnamefont {H.~E.}\ \bibnamefont {T\"ureci}},\
  }\href@noop {} {\bibfield  {journal} {\bibinfo  {journal} {Phys. Rev. Lett.}\
  }\textbf {\bibinfo {volume} {109}},\ \bibinfo {pages} {053601} (\bibinfo
  {year} {2012})}\BibitemShut {NoStop}%
\bibitem [{\citenamefont {Chin}\ \emph {et~al.}(2011)\citenamefont {Chin},
  \citenamefont {Prior}, \citenamefont {Huelga},\ and\ \citenamefont
  {Plenio}}]{Plenio_prl11}%
  \BibitemOpen
  \bibfield  {author} {\bibinfo {author} {\bibfnamefont {A.~W.}\ \bibnamefont
  {Chin}}, \bibinfo {author} {\bibfnamefont {J.}~\bibnamefont {Prior}},
  \bibinfo {author} {\bibfnamefont {S.~F.}\ \bibnamefont {Huelga}}, \ and\
  \bibinfo {author} {\bibfnamefont {M.~B.}\ \bibnamefont {Plenio}},\ }\href
  {\doibase 10.1103/PhysRevLett.107.160601} {\bibfield  {journal} {\bibinfo
  {journal} {Phys. Rev. Lett.}\ }\textbf {\bibinfo {volume} {107}},\ \bibinfo
  {pages} {160601} (\bibinfo {year} {2011})}\BibitemShut {NoStop}%
\bibitem [{\citenamefont {Bera}\ \emph {et~al.}(2014)\citenamefont {Bera},
  \citenamefont {Florens}, \citenamefont {Baranger}, \citenamefont {Roch},
  \citenamefont {Nazir},\ and\ \citenamefont {Chin}}]{BeraEtAlPRB14}%
  \BibitemOpen
  \bibfield  {author} {\bibinfo {author} {\bibfnamefont {S.}~\bibnamefont
  {Bera}}, \bibinfo {author} {\bibfnamefont {S.}~\bibnamefont {Florens}},
  \bibinfo {author} {\bibfnamefont {H.~U.}\ \bibnamefont {Baranger}}, \bibinfo
  {author} {\bibfnamefont {N.}~\bibnamefont {Roch}}, \bibinfo {author}
  {\bibfnamefont {A.}~\bibnamefont {Nazir}}, \ and\ \bibinfo {author}
  {\bibfnamefont {A.~W.}\ \bibnamefont {Chin}},\ }\href@noop {} {\bibfield
  {journal} {\bibinfo  {journal} {Phys. Rev. B}\ }\textbf {\bibinfo {volume}
  {89}},\ \bibinfo {pages} {121108} (\bibinfo {year} {2014})}\BibitemShut
  {NoStop}%
\bibitem [{\citenamefont {Vorrath}\ and\ \citenamefont
  {Brandes}(2005)}]{VorrathBrandesPRL05}%
  \BibitemOpen
  \bibfield  {author} {\bibinfo {author} {\bibfnamefont {T.}~\bibnamefont
  {Vorrath}}\ and\ \bibinfo {author} {\bibfnamefont {T.}~\bibnamefont
  {Brandes}},\ }\href@noop {} {\bibfield  {journal} {\bibinfo  {journal} {Phys.
  Rev. Lett.}\ }\textbf {\bibinfo {volume} {95}},\ \bibinfo {pages} {070402}
  (\bibinfo {year} {2005})}\BibitemShut {NoStop}%
\bibitem [{\citenamefont {Anders}(2008)}]{AndersNJP08}%
  \BibitemOpen
  \bibfield  {author} {\bibinfo {author} {\bibfnamefont {F.~B.}\ \bibnamefont
  {Anders}},\ }\href@noop {} {\bibfield  {journal} {\bibinfo  {journal} {New
  Journal of Physics}\ }\textbf {\bibinfo {volume} {10}},\ \bibinfo {pages}
  {115007} (\bibinfo {year} {2008})}\BibitemShut {NoStop}%
\bibitem [{\citenamefont {Winter}\ and\ \citenamefont
  {Rieger}(2014)}]{WinterRiegerPRB14}%
  \BibitemOpen
  \bibfield  {author} {\bibinfo {author} {\bibfnamefont {A.}~\bibnamefont
  {Winter}}\ and\ \bibinfo {author} {\bibfnamefont {H.}~\bibnamefont
  {Rieger}},\ }\href@noop {} {\bibfield  {journal} {\bibinfo  {journal} {Phys.
  Rev. B}\ }\textbf {\bibinfo {volume} {90}},\ \bibinfo {pages} {224401}
  (\bibinfo {year} {2014})}\BibitemShut {NoStop}%
\bibitem [{\citenamefont {Henriet}\ and\ \citenamefont
  {Le~Hur}(2016)}]{HenrietLeHurPRB16}%
  \BibitemOpen
  \bibfield  {author} {\bibinfo {author} {\bibfnamefont {L.}~\bibnamefont
  {Henriet}}\ and\ \bibinfo {author} {\bibfnamefont {K.}~\bibnamefont
  {Le~Hur}},\ }\href {\doibase 10.1103/PhysRevB.93.064411} {\bibfield
  {journal} {\bibinfo  {journal} {Phys. Rev. B}\ }\textbf {\bibinfo {volume}
  {93}},\ \bibinfo {pages} {064411} (\bibinfo {year} {2016})}\BibitemShut
  {NoStop}%
\bibitem [{\citenamefont {Nagy}\ and\ \citenamefont
  {Domokos}(2015)}]{NagyDomokosPRL15}%
  \BibitemOpen
  \bibfield  {author} {\bibinfo {author} {\bibfnamefont {D.}~\bibnamefont
  {Nagy}}\ and\ \bibinfo {author} {\bibfnamefont {P.}~\bibnamefont {Domokos}},\
  }\href@noop {} {\bibfield  {journal} {\bibinfo  {journal} {Phys. Rev. Lett.}\
  }\textbf {\bibinfo {volume} {115}},\ \bibinfo {pages} {043601} (\bibinfo
  {year} {2015})}\BibitemShut {NoStop}%
\bibitem [{\citenamefont {{Nagy}}\ and\ \citenamefont
  {{Domokos}}(2016)}]{NagyDomokos_arxiv16}%
  \BibitemOpen
  \bibfield  {author} {\bibinfo {author} {\bibfnamefont {D.}~\bibnamefont
  {{Nagy}}}\ and\ \bibinfo {author} {\bibfnamefont {P.}~\bibnamefont
  {{Domokos}}},\ }\href@noop {} {\bibfield  {journal} {\bibinfo  {journal}
  {ArXiv e-prints}\ } (\bibinfo {year} {2016})},\ \Eprint
  {http://arxiv.org/abs/1610.01124} {arXiv:1610.01124 [quant-ph]} \BibitemShut
  {NoStop}%
\bibitem [{\citenamefont {Liu}\ \emph {et~al.}(2011)\citenamefont {Liu},
  \citenamefont {Li}, \citenamefont {Huang}, \citenamefont {Li}, \citenamefont
  {Guo}, \citenamefont {Laine}, \citenamefont {Breuer},\ and\ \citenamefont
  {Piilo}}]{LiuEtAlNatPhys11}%
  \BibitemOpen
  \bibfield  {author} {\bibinfo {author} {\bibfnamefont {B.-H.}\ \bibnamefont
  {Liu}}, \bibinfo {author} {\bibfnamefont {L.}~\bibnamefont {Li}}, \bibinfo
  {author} {\bibfnamefont {Y.-F.}\ \bibnamefont {Huang}}, \bibinfo {author}
  {\bibfnamefont {C.-F.}\ \bibnamefont {Li}}, \bibinfo {author} {\bibfnamefont
  {G.-C.}\ \bibnamefont {Guo}}, \bibinfo {author} {\bibfnamefont {E.-M.}\
  \bibnamefont {Laine}}, \bibinfo {author} {\bibfnamefont {H.-P.}\ \bibnamefont
  {Breuer}}, \ and\ \bibinfo {author} {\bibfnamefont {J.}~\bibnamefont
  {Piilo}},\ }\href {http://dx.doi.org/10.1038/nphys2085} {\bibfield  {journal}
  {\bibinfo  {journal} {Nat Phys}\ }\textbf {\bibinfo {volume} {7}},\ \bibinfo
  {pages} {931} (\bibinfo {year} {2011})}\BibitemShut {NoStop}%
\bibitem [{\citenamefont {Sundaresan}\ \emph {et~al.}(2015)\citenamefont
  {Sundaresan}, \citenamefont {Liu}, \citenamefont {Sadri}, \citenamefont
  {Sz\ifmmode~\mbox{\H{o}}\else \H{o}\fi{}cs}, \citenamefont {Underwood},
  \citenamefont {Malekakhlagh}, \citenamefont {T\"ureci},\ and\ \citenamefont
  {Houck}}]{SundaresanEtAlPRX15}%
  \BibitemOpen
  \bibfield  {author} {\bibinfo {author} {\bibfnamefont {N.~M.}\ \bibnamefont
  {Sundaresan}}, \bibinfo {author} {\bibfnamefont {Y.}~\bibnamefont {Liu}},
  \bibinfo {author} {\bibfnamefont {D.}~\bibnamefont {Sadri}}, \bibinfo
  {author} {\bibfnamefont {L.~J.}\ \bibnamefont {Sz\ifmmode~\mbox{\H{o}}\else
  \H{o}\fi{}cs}}, \bibinfo {author} {\bibfnamefont {D.~L.}\ \bibnamefont
  {Underwood}}, \bibinfo {author} {\bibfnamefont {M.}~\bibnamefont
  {Malekakhlagh}}, \bibinfo {author} {\bibfnamefont {H.~E.}\ \bibnamefont
  {T\"ureci}}, \ and\ \bibinfo {author} {\bibfnamefont {A.~A.}\ \bibnamefont
  {Houck}},\ }\href@noop {} {\bibfield  {journal} {\bibinfo  {journal} {Phys.
  Rev. X}\ }\textbf {\bibinfo {volume} {5}},\ \bibinfo {pages} {021035}
  (\bibinfo {year} {2015})}\BibitemShut {NoStop}%
\bibitem [{\citenamefont {{Haeberlein}}\ \emph {et~al.}(2015)\citenamefont
  {{Haeberlein}}, \citenamefont {{Deppe}}, \citenamefont {{Kurcz}},
  \citenamefont {{Goetz}}, \citenamefont {{Baust}}, \citenamefont {{Eder}},
  \citenamefont {{Fedorov}}, \citenamefont {{Fischer}}, \citenamefont
  {{Menzel}}, \citenamefont {{Schwarz}}, \citenamefont {{Wulschner}},
  \citenamefont {{Xie}}, \citenamefont {{Zhong}}, \citenamefont {{Solano}},
  \citenamefont {{Marx}}, \citenamefont {{Garc{\'{\i}}a-Ripoll}},\ and\
  \citenamefont {{Gross}}}]{HaeberleinEtAl_arxiv15}%
  \BibitemOpen
  \bibfield  {author} {\bibinfo {author} {\bibfnamefont {M.}~\bibnamefont
  {{Haeberlein}}}, \bibinfo {author} {\bibfnamefont {F.}~\bibnamefont
  {{Deppe}}}, \bibinfo {author} {\bibfnamefont {A.}~\bibnamefont {{Kurcz}}},
  \bibinfo {author} {\bibfnamefont {J.}~\bibnamefont {{Goetz}}}, \bibinfo
  {author} {\bibfnamefont {A.}~\bibnamefont {{Baust}}}, \bibinfo {author}
  {\bibfnamefont {P.}~\bibnamefont {{Eder}}}, \bibinfo {author} {\bibfnamefont
  {K.}~\bibnamefont {{Fedorov}}}, \bibinfo {author} {\bibfnamefont
  {M.}~\bibnamefont {{Fischer}}}, \bibinfo {author} {\bibfnamefont {E.~P.}\
  \bibnamefont {{Menzel}}}, \bibinfo {author} {\bibfnamefont {M.~J.}\
  \bibnamefont {{Schwarz}}}, \bibinfo {author} {\bibfnamefont {F.}~\bibnamefont
  {{Wulschner}}}, \bibinfo {author} {\bibfnamefont {E.}~\bibnamefont {{Xie}}},
  \bibinfo {author} {\bibfnamefont {L.}~\bibnamefont {{Zhong}}}, \bibinfo
  {author} {\bibfnamefont {E.}~\bibnamefont {{Solano}}}, \bibinfo {author}
  {\bibfnamefont {A.}~\bibnamefont {{Marx}}}, \bibinfo {author} {\bibfnamefont
  {J.-J.}\ \bibnamefont {{Garc{\'{\i}}a-Ripoll}}}, \ and\ \bibinfo {author}
  {\bibfnamefont {R.}~\bibnamefont {{Gross}}},\ }\href@noop {} {\bibfield
  {journal} {\bibinfo  {journal} {ArXiv e-prints}\ } (\bibinfo {year}
  {2015})},\ \Eprint {http://arxiv.org/abs/1506.09114} {arXiv:1506.09114
  [cond-mat.mes-hall]} \BibitemShut {NoStop}%
\bibitem [{\citenamefont {Liu}\ and\ \citenamefont
  {Houck}(2016)}]{LiuHouckNatPhys16}%
  \BibitemOpen
  \bibfield  {author} {\bibinfo {author} {\bibfnamefont {Y.}~\bibnamefont
  {Liu}}\ and\ \bibinfo {author} {\bibfnamefont {A.~A.}\ \bibnamefont
  {Houck}},\ }\href {http://dx.doi.org/10.1038/nphys3834} {\bibfield  {journal}
  {\bibinfo  {journal} {Nat Phys}\ }\textbf {\bibinfo {volume} {advance online
  publication}},\  (\bibinfo {year} {2016})}\BibitemShut {NoStop}%
\bibitem [{\citenamefont {Fink}\ \emph {et~al.}(2009)\citenamefont {Fink},
  \citenamefont {Bianchetti}, \citenamefont {Baur}, \citenamefont {G\"oppl},
  \citenamefont {Steffen}, \citenamefont {Filipp}, \citenamefont {Leek},
  \citenamefont {Blais},\ and\ \citenamefont {Wallraff}}]{FinkEtAlPRL09}%
  \BibitemOpen
  \bibfield  {author} {\bibinfo {author} {\bibfnamefont {J.~M.}\ \bibnamefont
  {Fink}}, \bibinfo {author} {\bibfnamefont {R.}~\bibnamefont {Bianchetti}},
  \bibinfo {author} {\bibfnamefont {M.}~\bibnamefont {Baur}}, \bibinfo {author}
  {\bibfnamefont {M.}~\bibnamefont {G\"oppl}}, \bibinfo {author} {\bibfnamefont
  {L.}~\bibnamefont {Steffen}}, \bibinfo {author} {\bibfnamefont
  {S.}~\bibnamefont {Filipp}}, \bibinfo {author} {\bibfnamefont {P.~J.}\
  \bibnamefont {Leek}}, \bibinfo {author} {\bibfnamefont {A.}~\bibnamefont
  {Blais}}, \ and\ \bibinfo {author} {\bibfnamefont {A.}~\bibnamefont
  {Wallraff}},\ }\href {\doibase 10.1103/PhysRevLett.103.083601} {\bibfield
  {journal} {\bibinfo  {journal} {Phys. Rev. Lett.}\ }\textbf {\bibinfo
  {volume} {103}},\ \bibinfo {pages} {083601} (\bibinfo {year}
  {2009})}\BibitemShut {NoStop}%
\bibitem [{\citenamefont {Grezes}\ \emph {et~al.}(2014)\citenamefont {Grezes},
  \citenamefont {Julsgaard}, \citenamefont {Kubo}, \citenamefont {Stern},
  \citenamefont {Umeda}, \citenamefont {Isoya}, \citenamefont {Sumiya},
  \citenamefont {Abe}, \citenamefont {Onoda}, \citenamefont {Ohshima},
  \citenamefont {Jacques}, \citenamefont {Esteve}, \citenamefont {Vion},
  \citenamefont {Esteve}, \citenamefont {M\o{}lmer},\ and\ \citenamefont
  {Bertet}}]{GrezesEtAlPRX14}%
  \BibitemOpen
  \bibfield  {author} {\bibinfo {author} {\bibfnamefont {C.}~\bibnamefont
  {Grezes}}, \bibinfo {author} {\bibfnamefont {B.}~\bibnamefont {Julsgaard}},
  \bibinfo {author} {\bibfnamefont {Y.}~\bibnamefont {Kubo}}, \bibinfo {author}
  {\bibfnamefont {M.}~\bibnamefont {Stern}}, \bibinfo {author} {\bibfnamefont
  {T.}~\bibnamefont {Umeda}}, \bibinfo {author} {\bibfnamefont
  {J.}~\bibnamefont {Isoya}}, \bibinfo {author} {\bibfnamefont
  {H.}~\bibnamefont {Sumiya}}, \bibinfo {author} {\bibfnamefont
  {H.}~\bibnamefont {Abe}}, \bibinfo {author} {\bibfnamefont {S.}~\bibnamefont
  {Onoda}}, \bibinfo {author} {\bibfnamefont {T.}~\bibnamefont {Ohshima}},
  \bibinfo {author} {\bibfnamefont {V.}~\bibnamefont {Jacques}}, \bibinfo
  {author} {\bibfnamefont {J.}~\bibnamefont {Esteve}}, \bibinfo {author}
  {\bibfnamefont {D.}~\bibnamefont {Vion}}, \bibinfo {author} {\bibfnamefont
  {D.}~\bibnamefont {Esteve}}, \bibinfo {author} {\bibfnamefont
  {K.}~\bibnamefont {M\o{}lmer}}, \ and\ \bibinfo {author} {\bibfnamefont
  {P.}~\bibnamefont {Bertet}},\ }\href {\doibase 10.1103/PhysRevX.4.021049}
  {\bibfield  {journal} {\bibinfo  {journal} {Phys. Rev. X}\ }\textbf {\bibinfo
  {volume} {4}},\ \bibinfo {pages} {021049} (\bibinfo {year}
  {2014})}\BibitemShut {NoStop}%
\bibitem [{\citenamefont {Kakuyanagi}\ \emph {et~al.}(2016)\citenamefont
  {Kakuyanagi}, \citenamefont {Matsuzaki}, \citenamefont {D\'eprez},
  \citenamefont {Toida}, \citenamefont {Semba}, \citenamefont {Yamaguchi},
  \citenamefont {Munro},\ and\ \citenamefont {Saito}}]{KakuyanagiEtAlPRL16}%
  \BibitemOpen
  \bibfield  {author} {\bibinfo {author} {\bibfnamefont {K.}~\bibnamefont
  {Kakuyanagi}}, \bibinfo {author} {\bibfnamefont {Y.}~\bibnamefont
  {Matsuzaki}}, \bibinfo {author} {\bibfnamefont {C.}~\bibnamefont {D\'eprez}},
  \bibinfo {author} {\bibfnamefont {H.}~\bibnamefont {Toida}}, \bibinfo
  {author} {\bibfnamefont {K.}~\bibnamefont {Semba}}, \bibinfo {author}
  {\bibfnamefont {H.}~\bibnamefont {Yamaguchi}}, \bibinfo {author}
  {\bibfnamefont {W.~J.}\ \bibnamefont {Munro}}, \ and\ \bibinfo {author}
  {\bibfnamefont {S.}~\bibnamefont {Saito}},\ }\href {\doibase
  10.1103/PhysRevLett.117.210503} {\bibfield  {journal} {\bibinfo  {journal}
  {Phys. Rev. Lett.}\ }\textbf {\bibinfo {volume} {117}},\ \bibinfo {pages}
  {210503} (\bibinfo {year} {2016})}\BibitemShut {NoStop}%
\end{thebibliography}
%

\end{document}


\title{Supplementary Information for "Dissipation-Induced Superradiance in a Non-Markovian Open Dicke Model"}
 \author{Orazio Scarlatella}
\author{Marco Schir\'o}
\affiliation{Institut de Physique Th\'{e}orique, Universit\'{e} Paris Saclay, CNRS, CEA, F-91191 Gif-sur-Yvette, France}
\date{\today}

\date{\today}
\pacs{42.65.Sf,42.50.pq,05.70.Ln}
\maketitle

This Supplementary Information is organized as follows: in the first section we solve the Dicke model in mean field recovering the temperature dependent phase diagram; then we focus on the Dicke model coupled to a bath obtaining, in the second section, an effective Keldysh action in the case of a frequency dependent bath. The third section discusses the case of a Markovian bath within the Lindblad master equation, while the last section the mapping to an effective spin-boson model.

\section{Equilibrium Finite Temperature Solution of Dicke Model}

Here we briefly recall the equilibrium solution of the closed-Dicke model with a mean field decoupling of the light-matter interaction, which neglects some correlation between the cavity mode and the large spin but naturally allows to describe the finite temperature phase transition. We start from the Dicke Hamiltonian, Eq.~(1), of the main text and decouple the light-matter interaction $H_{int}=\frac{\lambda}{\sqrt{N}}\left(a+a^{\dagger}\right)\sum_i S^x_i$ as
\be
H_{int}\simeq 
\frac{\lambda}{\sqrt{N}}\left(a+a^{\dagger}\right)\sum_i \langle S^x_i\rangle +
\frac{\lambda}{\sqrt{N}}\langle a+a^{\dagger}\rangle
\sum_i \langle S^x_i\rangle
\ee
to obtain two effective problems describing a displaced harmonic oscillator  
\be
 H_{ph} = \omega_0 a^{\dagger}a+\eta(\beta)\left(a+a^{\dagger}\right)
\ee
and a collection of independent spins in presence of a trasverse feld
\be 
H_{spin} = \omega_q \sum_i S_i^z + \zeta(\beta) \sum_i S_i^x
\ee
The photon and spin sectors are coupled only through the self-consistency conditions
\bea
 \eta(\beta) = \frac{\lambda}{\sqrt{N}} \sum_i \langle  S^x_i \rangle_{spin}\\
 \zeta(\beta) = \frac{\lambda}{\sqrt{N}} \langle x \rangle_{ph} 
\eea
By computing the average spin polarization $\langle S^x_i\rangle $ at fixed $\zeta$ as well as the average photon displacement $\langle x\rangle$ at fixed $\eta$, and plugging these into the self-consistency conditions above, one can obtain two coupled equations. We solve for the photon displacement $\langle x\rangle$, the result for the spin can be found in the same way. The self-consistent equation admits  two solutions, a symmetric one, $\langle x\rangle=0$,  corresponding to the normal phase, as well as a non trivial one which can be obtained solving
\begin{align}
z &= \tanh{\left(\beta \frac{2 \lambda^2}{ \omega_0}z\right)},& 
z &=\frac{\omega_0}{2 \lambda^2} \sqrt{\frac{\omega_q^2}{4} + \frac{\lambda^2}{N} \langle  x \rangle^2}. \label{eq:mean_field_eq}
\end{align}
At zero temperature, $\beta=\infty$, this identifies a quantum phase transition at the critical coupling $\lambda_{c} ^ 2 =\frac{\omega_0 \omega_q}{4}$, as obtained in~\cite{EmaryBrandesPRE03}. We notice that the mean field decoupling gives a very poor description of the normal phase, since the coupling between photon and spin only happens in this description through a finite order parameter. However this approach allows to capture the finite temperature transition from the super-radiant to the normal phase above a critical temperature \begin{align}
T_c = \frac{\omega_q}{2 \tanh^{-1}{\left( \frac{\lambda_{c} ^2}{\lambda^2} \right)}}.
\end{align}
as one can obtain from Eq.~(\ref{eq:mean_field_eq}). We plot this critical temperature as a function of light-matter coupling in the main text, figure 1.
\section{Effective Keldysh Action and Cavity Photon Propagator}

We now consider the Dicke Hamiltonian coupled to a bath and discuss the derivation of Keldysh effective action for the photon and qubit. We start from the Hamiltonian of the problem, after writing the large spin in terms of a bosonic mode, which reads as $H=H_0+H_{bath}+H_I$ with
\begin{align}
\label{eq:bathFullHamiltonian}
H_0&= \omega_0 a^\dagger a + \omega_q b^\dagger b + \lambda (a^\dagger + a)(b^\dagger + b ) \\
H_{bath} &= \sum \omega_k b_k^\dagger b_k\;\qquad
H_I =  \sum_k g_k (a^\dagger + a )( b_k^\dagger +b_k) \nonumber
\end{align}
The associated Keldysh action written in terms of the bosonic displacements $x=a^{\dagger}+a$, $y=b^\dagger + b $, $z_k= b^\dagger_k + b_k$ and in the classical-quantum basis~\cite{KamenevBook} reads $S=S_0+S_{bath}+S_I$ where
\begin{widetext}
\begin{gather*}
\begin{split}
S_0 &= \int_{-\infty}^\infty dt \: dt' \: \frac{1}{2} \left[ \begin{pmatrix}
x_c (t) & x_q (t) 
\end{pmatrix} \begin{pmatrix}
0 & [\chi_0^{-1}]^A \\ [\chi_0^{-1}]^R & [\chi_0^{-1}]^K 
\end{pmatrix}_{t-t'} \begin{pmatrix}
 x_c(t') \\ x_q(t') 
\end{pmatrix} +  \begin{pmatrix}
y_c (t) & y_q (t) 
\end{pmatrix} \begin{pmatrix}
0 & [\zeta^{-1}]^A \\ [\zeta^{-1}]^R & [\zeta^{-1}]^K 
\end{pmatrix}_{t-t'} \begin{pmatrix}
 y_c(t') \\ y_q(t') 
\end{pmatrix} \right]+ 
\\ &
+\lambda\int_{-\infty}^\infty dt 
\left[   
\begin{pmatrix}
x_c (t) & x_q (t) 
\end{pmatrix} \tau_x \begin{pmatrix}
y_c (t) \\ y_q (t) 
\end{pmatrix} 
\right]
\end{split}
\end{gather*}
while 
\begin{eqnarray}
S_{bath}=
\frac{1}{2}\int_{-\infty}^\infty dt dt'\left[\sum_k  \begin{pmatrix}
z_{c,k} (t) & z_{q,k} (t) 
\end{pmatrix} \begin{pmatrix}
0 & [\zeta_k^{-1}]^A \\ [\zeta_k^{-1}]^R & [\zeta_k^{-1}]^K 
\end{pmatrix}_{t-t'} \begin{pmatrix}
 z_{c,k}(t') \\ z_{q,k}(t') 
\end{pmatrix} \right]\\
S_I= \lambda\int_{-\infty}^\infty dt \left[   
 \begin{pmatrix}
x_c (t) & x_q (t) 
\end{pmatrix} \tau_x \sum_k g_k \begin{pmatrix}
z_{c,k} (t) \\ z_{q,k} (t) 
\end{pmatrix}  \right]
\end{eqnarray}
In the above equations, we have introduced the Pauli matrix $\tau_x = \begin{pmatrix}
0 & 1 \\ 1 & 0 
\end{pmatrix}$ while $\chi_0$, $\zeta$, $\zeta_k$ are the non-interacting Green's functions of bosonic displacements, i.e. for example 
\bea
\zeta^{R}_k(t-t')\ =-i\theta(t-t')\langle\left[z_k(t-t'),z_k(0)\right]\rangle_{bath}\\
\zeta^{A}_k(t-t')\ =i\theta(t'-t)\langle\left[z_k(t-t'),z_k(0)\right]\rangle_{bath}\\
\zeta^{K}_k(t-t')\ =-i\langle \left\{z_k(t-t'),z_k(0)\right\}\rangle_{bath}
\eea
\end{widetext}
with similar expressions for $\chi_0$ and $\zeta$. We then integrate out the bath degrees of freedom $\zeta_k$ which can be done exactly since the action is gaussian. The effect of the bath is to generate a term in the action which is non-local in time and accounts for dissipation, which reads
\small
\begin{equation}\label{eqn:SeffK}
S_{eff} = S_0
-  \int_{-\infty}^\infty dt \: dt' \begin{pmatrix}
x_c (t) & x_q (t) 
\end{pmatrix} \hat{\Delta}(t-t')
\begin{pmatrix}
 x_c(t') \\ x_q(t') 
\end{pmatrix}
\end{equation}
\normalsize
where the bath-induced self-energy reads
\begin{align*}
\hat{\Delta}(t-t') &= 
\begin{pmatrix}
0 & \Delta^A \\  \Delta^R & \Delta^K
\end{pmatrix}_{t-t'}  =
\sum_k g_k^2 \sigma_x \begin{pmatrix}
\zeta_k^K & \zeta_k^R \\ \zeta_k^A & 0 
\end{pmatrix}_{t-t'} \sigma_x
\end{align*}
From the effective action in Eq.~(\ref{eqn:SeffK}) we can obtain easily the cavity photon propagator $\chi(t-t')$ defined as
\begin{eqnarray}
\begin{split}
\hat{\chi}(t-t') &=   \begin{pmatrix}
\chi^K & \chi^R \\ \chi^A & 0 
\end{pmatrix}_{t-t'} = \\
=&\int \mathcal{D}[x,y]  \begin{pmatrix}
x_c(t) \\ x_q(t) 
\end{pmatrix} \begin{pmatrix}
x_c(t) & x_q(t) 
\end{pmatrix} e^{i S_{eff}}
\end{split}
\end{eqnarray} 
We first perform the gaussian integral on the other bosonic mode $y_{c/q}$ which results in a further self-energy correction due to light matter interaction
\begin{align*}
\hat{\Sigma}_\lambda(t-t') &= 
\begin{pmatrix}
0 & \Sigma^A \\  \Sigma^R & \Sigma^K
\end{pmatrix}_{t-t'} 
&= \lambda^2 \tau_x \begin{pmatrix}
\zeta^K & \zeta^R \\ \zeta^A & 0
\end{pmatrix}_{t-t'}  \tau_x \\
\end{align*}
Then from this result we can immediately read out the inverse susceptibility which takes the form 
\be
\label{eq:inverseSusceptibility}
\hat{\chi}^{-1} = \hat{\chi}_0^{-1} - \hat{\Sigma}_\lambda - \hat{\Delta}
\ee
and whose retarded component is given in the main text.

\section{Dicke Model coupled to a Markovian Bath}

The open Dicke model coupled to a Markovian bath has been studied previously by a number of authors (see Refs [16-25] in the main text) so in this section we only briefly recall the results of the analysis obtained within Master Equation or Keldysh Field Theory approaches, by emphasizing in particular the role of frequency renormalization due to the bath and the essence of Markov approximation.

When deriving the Lindblad master equation~\cite{breuerPetruccione}  one typically starts from the system plus bath Hamiltonian, that for simplicity we write here in the rotating wave approximation (RWA) with respect to Eq.~(\ref{eq:bathFullHamiltonian}), i.e.
\be 
H=H_0 + H_{bath}+ \sum_k g_k \left(a^\dagger b_k+ b_k^\dagger a\right)
\ee
then trace out the degrees of freedom of the environment within second order (Born) perturbation theory and perform the Markov approximation to obtain a local-in-time evolution for the system density matrix
\begin{equation}
\partial_t \rho = -i [\tilde{H}_D,\rho] 
+\mathcal{L}(\rho). \label{eq:Born-MarkovEquation}
\end{equation}
where the Lindblad dissipator is defined by
\begin{gather}
\begin{split}
\mathcal{L}(\rho) &= \kappa [n(\omega_0)+1] (2 a \rho a^\dagger -\{a^\dagger a , \rho \} ) + \\ &+ \kappa n(\omega_0) (2 a^\dagger \rho a-\{a a^\dagger , \rho \} ),
\end{split} \label{eq:lindbladDissipator} 
\end{gather}
where $n(\omega_0)$ is the Bose-Einstein distribution, the loss rate $\kappa = J(\omega_0)$ and the bath spectral function is $J(\omega) = \pi \sum_k g_k^2 \delta(\omega-\omega_k)$.  The standard manipulations that bring to Eq.~(\ref{eq:lindbladDissipator}) usually do not properly take into account light-matter interaction, but rather assume the photon subsystem to evolve freely. A proper treatment would be in principle required in the so called ultrastrong coupling regime~\cite{CiutiCarusottoPRA06,beaudoin_dissipation_2011}.
\begin{figure}[t]
\begin{center}
\epsfig{figure=../Plot/lambshift.pdf,scale=0.33}
\caption{Markovian Open Dicke Model - Critical bath coupling for the SR transition as a function of light-matter interaction. We compare the result obtained without (dashed line) and with the bath-induced frequency renormalization (Lamb Shift), the latter for different bath exponents $s$. }
\label{fig:fig1SM}
\end{center}
\end{figure}
The unitary part of the evolution is described by the Dicke Hamiltonian $\tilde{H}_D$, where the frequency of the cavity mode is renormalized by the bath
\be\label{eqn:omegaRWA}
\tilde{\omega}_0= \omega_0+P\int_{0}^{\infty} \frac{d \xi}{\pi} \frac{J(\xi)}{\omega_0 - \xi}\equiv
\omega_0-c\kappa
\ee
where $c=-(1/2\pi)P\int_0^1 dx x^s/(\omega_0/\omega_c-x)$ depends on $\omega_0/\omega_c$ and the bath exponent $s$. This frequency renormalization has a similar structure with respect to the one discussed in the main text, beside (i) the different prefactor coming from the inclusion of counter-rotating terms in the full Keldysh calculation and (ii) the fact that within the Lindblad master equation the real-part of the bath self-energy is sampled at the bare photon frequency. In terms of the renormalized Dicke Hamiltonian $\tilde{H}_D$ one can proceed as usual and write down equation of motions for the average values $ \langle a \rangle $, $ \langle S^\pm \rangle $ and $ \langle S^z \rangle $ 
\begin{subequations}\label{eq:systEqOfMotLind}
\begin{align}
\partial_t  \langle a \rangle  &= - i \tilde{\omega}_0 \langle a \rangle  - \kappa  \langle a \rangle  - \frac{i \lambda}{\sqrt{N}}   \langle S^+ + S^- \rangle , \\
\partial_t  \langle S^+ \rangle  &= i \omega_q  \langle S^+ \rangle   - \frac{2i \lambda}{\sqrt{N}}  \langle  S^z (a^\dagger + a )  \rangle , \\
\partial_t  \langle S^z \rangle  &= -\frac{i \lambda}{\sqrt{N}}  \langle (a^\dagger + a) ( S^+ - S^-)  \rangle .
\end{align}
\end{subequations}
These can be solved with a mean field decoupling, using the conservation of total spin $S^2$, to obtain the stationary state solution for the cavity mode order parameter $\langle x\rangle=\langle a+a^{\dagger}\rangle$
\begin{align*}
 \langle x  \rangle  &= \frac{4 \lambda}{\Omega_0}\sqrt{\frac{N}{4}\left(1-\frac{\lambda_c^4}{\lambda^4} \right)}, & \Omega_0 &= \frac{(\tilde{\omega}_0)^2 + \kappa^2}{\tilde{\omega}_0}
\end{align*}
 which becomes non-zero above a critical coupling
\begin{gather}
\tilde{\lambda}_c^M (\kappa)  = \sqrt{((\tilde{\omega}_0)^2+\kappa^2)\omega_q/4 \tilde{\omega}_0}.
\label{eqn:lacbathlindRWA} 
\end{gather}

If we disregard the frequency renormalization coming from the bath modes, Eq.~(\ref{eqn:omegaRWA}),  we recover the well-known result for a Lindblad-Markovian bath, 
\be\label{eqn:lacbathlind}
\lambda_c^M (\kappa)  = \sqrt{ (\omega_0^2+\kappa^2)\omega_q/4 \omega_0}
\ee
otherwise the two results differ for the bath-induced frequency renormalization, Eq.~(\ref{eqn:omegaRWA}), which is linear in the dissipation strenght.


In figure~\ref{fig:fig1SM}  we plot the critical dissipation strenght $\kappa_c(\lambda)$, obtained from Eqs.~(\ref{eqn:lacbathlindRWA},\ref{eqn:lacbathlind}) for different values of the bath exponent $s$. 
We see that upon including the frequency renormalization due to the bath, the shape of the phase boundary at small dissipation changes, an effect which is more pronounced for sub-ohmic environments consistently with their abundance of low-frequency modes.  The phase boundary approaches, at least qualitatively, the behavior seen in the main text for a generic, and possibly non-markovian, environment $J(\omega)$. 
This means that, at least for weak coupling to the bath, there is not much of a difference between the two kind of environments as far as the phase transition is concerned. At larger dissipation instead, qualitative deviations appear. These can be explained by noticing that the master equation result in Eq.~(\ref{eqn:lacbathlindRWA}) can be recovered starting from the full Keldysh solution for a system-bath coupling in the RWA and by making the effective action local in time, the hallmark of Markovian systems~\cite{torre_keldysh_2013,BuchholdEtAlPRA13}. This amounts to evaluate the real and imaginary part of the bath self-energy at the photon frequency $\omega_0$, \emph{before} taking the zero frequency limit which is necessary to probe the phase transition. Instead, for a generic frequency dependent bath as in the main text case one can compute the cavity photon susceptibility and evaluate it at zero frequency for any give bath function, obtaining the result shown in Eq. (6) of the main text.

\section{Effective Spin Boson Model}

If we start from the full Dicke plus bath Hamiltonian, Eq. (1-2) in the main text, and proceed by integrating out all the bosonic modes, i.e. first the bath and then the cavity photon, we obtain an effective action for the spins degrees of freedom only which has the form
\be\label{eqn:Seff}
 \m{S}_{eff}= \frac{1}{N}\sum_{ij}\int d\tau d\tau' S^x_i(\tau)\,\m{J}_{eff}\left(\tau-\tau'\right)S^x_j(\tau')+\m{S}_{loc}
\ee
where $\m{S}_{loc}$ is the contribution from the atomic frequencies, $\omega_q\sum_i S^z_i$, which acts as a transverse field here, while the effective time-dependent exchange mediated by the cavity photon reads
\be
\m{J}_{eff}(\tau)=-2\lambda^2\langle T_{\tau} x(\tau)x(0)\rangle
\ee
with $x=a+a^{\dagger}$. This interaction is completely determined by the non-interacting, $\lambda=0$, photon Green's function and can be written in integral form as
\be\label{eqn:Jeff_tau}
 \m{J}_{eff}(\tau)=
 -\int_0^{\infty} \frac{d\omega}{\pi}\left(\frac{\cosh((\beta-2\tau) \omega/2)}{\sinh(\beta\omega/2)}\right) 
\m{J}_{eff}(\omega)
\ee
where 
\be
\m{J}_{eff}(\omega)=
\frac{4\lambda^2\omega_0^2 J(\omega)}{\left(\omega^2-\omega_0^2-2\omega_0\Lambda(\omega)\right)^2+(2\omega_0J(\omega))^2} 
\ee
Notice that this function at low frequency share the same power-law structure of the bath, $J(\omega)$, a fact that will play a crucial role in the following. To understand the physics encoded in the action~(\ref{eqn:Seff}) let's start from the case $N=1$, which describes a spin boson model in a renormalized bath. At low-frequency we can define indeed a renormalized dissipation strenght $\m{J}_{eff}(\omega\rightarrow0)=2\pi\alpha_{eff}\omega_c (\omega/\omega_c)^s$ which reads
\be
\alpha_{eff}=\frac{\lambda^2\kappa}{\pi \tilde{\omega}_0^2(\kappa)\omega_c} 
\ee
The $N=1$ spin-boson model has a localization transition upon increasing $\alpha_{eff}$ above a critical point. What happens if we further increase $N$? It is useful to rewrite the action as
\bea
 \m{S}_{eff}= \frac{1}{N}\sum_{i}\int d\tau d\tau' S^x_i(\tau)\,\m{J}_{eff}\left(\tau-\tau'\right)S^x_i(\tau')+ \m{S}_{loc}+
 \nonumber\\
 +\frac{1}{N}\sum_{i\neq j}\int d\tau d\tau' S^x_i(\tau)\,\m{J}_{eff}\left(\tau-\tau'\right)S^x_j(\tau')\nonumber
\eea
The first two terms describe a collection of independent spin-boson problems, with a bath which is suppressed by a factor $1/N$. The second term describes an all-to-all coupling between spin boson models at different sites. This term, differently than the first one, has a well defined $N\rightarrow\infty$ limit. Indeed if we take only the short-range (in time) part of the exchange and we write it as
\be
 \m{J}_{eff}(\tau)\sim \bar{\m{J}}_{eff}\delta(\tau) 
\ee
with 
\be\label{eqn:Jeff_mf}
\bar{\m{J}}_{eff}=\int d\tau \m{J}_{eff}(\tau)=\m{J}_{eff}(i\omega_n=0)=-\frac{4\lambda^2}{\tilde{\omega}_0(\kappa)}
\ee
the second term becomes local in time and describes a fully connected Ising model in a transverse field,
\be\label{eqn:Ising}
H_{eff} = \frac{\bar{\m{J}}_{eff}}{N}\sum_{ij}S^x_i S^x_j+\omega_q\sum_i S^z_i
\ee 
  which has, in the thermodynamic limit $N=\infty$, a quantum phase transition  when $\bar{\m{J}}_{eff}=-\omega_q$, a condition which immediately gives the mean field phase boundary in Eq.~(6) of the main text.  This condition translates onto a condition for the dissipation strength of an effective (mean-field) spin boson model with bath spectral function $J^{\infty}_{eff}(\omega)=2\pi\alpha^{\infty}_{eff}\omega_c(\omega/\omega_c)^s$. Indeed we can say that the static exchange $\bar{\m{J}}_{eff}=\int d\tau \m{J}_{eff}(\tau)$ corresponds to an effective dissipation $\alpha^{\infty}_{eff}$ such that Eq.~(\ref{eqn:Jeff_tau}) holds with $\m{J}_{eff}(\omega)\rightarrow J^{\infty}_{eff}(\omega)$, i.e.
\be
 \bar{\m{J}}_{eff}=-(2/\pi)\int d\omega \frac{J^{\infty}_{eff}(\omega)}{\omega}=-\frac{4\alpha^{\infty}_{eff}\omega_c}{s}
\ee 
 Using this result, together with Eq~(\ref{eqn:Jeff_mf}), we obtain a condition between $\alpha^{\infty}_{eff}$ and $\lambda,\kappa$, namely
 \be
 \alpha^{\infty}_{eff}=\frac{\lambda^2 s}{\omega_c\tilde{\omega}_0(\kappa)} 
 \ee
The above analysis suggests to define an $N$-dependent effective dissipation $\alpha_{eff}^N$ scaling as
\be
\alpha^N_{eff}=\alpha^{\infty}_{eff}+ \alpha_{eff}/N
\ee  
 In order to extract the scaling of the critical point $\lambda_c(\kappa)$ with $N$ we can compare this expression with the critical point $\alpha^N_c$ obtained from QMC simulations~\cite{WinterRiegerPRB14}, which gives the form
 \be
 \alpha^N_c=\alpha^{\infty}_c+\delta\alpha_c/N 
 \ee
 where $\alpha^{\infty}_c=\frac{s\omega_q}{4\omega_c} $ and $\delta\alpha_c$ independent from $s$. Notice the 
slightly different definition of various parameters in this work as compared to Ref.~\onlinecite{WinterRiegerPRB14}, which need to be taken into account when performing the comparison. From the above results we get 
\be
\lambda_c^N=\lambda^{\infty}_c+\frac{\delta\lambda}{N}
\ee
 with $\delta\lambda=\frac{1}{2}\left(\delta\alpha_c(\omega_q)/\alpha_c^{\infty}-\kappa/\pi s\tilde{\omega}_0(\kappa)\right)$, a quantity which can be positve or negative,  at fixed $\kappa$, depending on the frequencies $\omega_0,\omega_q$.


%